# Large Scale Structure[a]

## *The seventies & Forty years later: From Clusters to Clusters*

GUIDO CHINCARINI:
*University of Milano – Bicocca and Astronomical Observatory of Brera*
*Milano, Italy*

I describe the beginning, ~ 1970, of the spectroscopic redshift surveys and the discovery of the superclusters filamentary structures and voids. This changed the view of the distribution of luminous light from the way we knew it at the end of the sixties, a uniform distribution of galaxies with clusters superimposed, and the way we understood it during the seventies, clusters imbedded in filamentary structures. We planned in brief to understand the distribution of galaxies and the effect of their environment since to learn about the formation and evolution of the Universe, we need to know first how it is now. The very large surveys, first among these the CfA, started in the eighties thanks also to dedicated telescopes; during that period, the distribution of mass on large scales was also measured. A next step, in addition to the many and deep galaxies survey that are going on, could be a deep cluster survey to detect distant clusters, improve the accuracy of the clusters physical parameters as a function of redshift and have a more robust probe for Cosmology.

*Ego sum ergo Mundus est*
*Rio de Janeiro lecture*

**Prologue:**

I had the privilege to witness the development of the field, Large Scale Structure, from its very beginning (I contributed to the very initial phase of the field) to the most recent achievements. For this reason I have been invited to give a brief history, or better a personal view, about the initial steps of the study of the Large Scale Structure outlining the way I contributed to it. On the other hand it is more fun to emphasize also the great results of some of my peers touching upon a few fundamental achievements in the field. After doing this, I will conclude with a brief look to the future, where do we go from here? I am quite aware, however, that other scientists may have seen the evolving of the field in a different way. On the other hand some of the papers written in the past are misleading, or at least incomplete and do not reflect the early literature and achievements on this matter, this writing may perhaps clarify a few points.

The study of the Large Scale Structure of the Universe via spectroscopic surveys opened new ways. This is my view of how it begun.

In my letter to Nature (1978)[1] I give statistical evidence of the existence of large scale "holes", that is the existence of regions of the Cosmo in which we do not detect any galaxy, and I mentioned some of that paper's results at the Tallin meeting (see my comment following the paper by Tifft and Gregory)[2]. It also took a long time (about two years) to Rood and myself to write the Sky and Telescope article (1980)[3]. The writer, obviously, was Herb but we could not find the proper style till, at some point, Herb was satisfied and decided he had found the proper structure for the Sky and Telescope article. Rood and I discussed the draft for a long time at the University of Oklahoma and, at the time, we were very uncertain whether to use "holes" or "voids", a word used also by

---

[a] Invited talk at the 13th Marcel Grossman conference, Stockholm 2012.

Gregory and Thompson with the same meaning. We asked about this matter John Cowan (the matter was important because we were dealing with a magazine for amateurs and laymen) and John's verdict was:" Voids is a better English for what you mean" and that was it. We submitted the article shortly afterwards.

When Rood and I started to work in clusters of galaxies and later discovered Voids and Superclusters of galaxies in three dimensions (the beginning of the redshift surveys), I, and partly Herb, were rather ignorant about part of the past literature on the large scale distribution of galaxies and however I feel this helped in avoiding any bias and properly and naively approaching Nature. On the other hand I now refreshed the memory looking at the literature and discovering, as usual, few things I did not know at the time.

Thompson and Gregory did excellent and original work following a similar path Rood and I had started and the line set up by Bill Tifft. Einasto and his group developed on ideas that were going around with conclusions based on published, and however largely incomplete, data or catalogues. Before the meeting we had in Tallin, Einasto and his group were indeed concerned mainly with what they called "hypergalaxies" while after that meeting they got deeply involved in the discussion of super-clustering thanks also to the work carried out by Zeldovich, Doroshkevich and Shandarin. Indeed Zeldovich was very interested in our new findings based on redshift observations and also Gerard de Vaucouleurs considered it a break through in spite of his different conception of the Universe.

In the 5$^{th}$ section of this article I will try to jump from the old time activities to part of the fascinating accomplishments occurring today in the field. As usual the future, because of the unknown and the related feeling of discovery, given the tools at hand, is even more fascinating of the past. The accurate work generally leads to new wonders.

As I mentioned earlier this paper is in large part a personal recollection and it is not meant at all to review[b] the field. Among the many meetings on Cosmology I refer to that held in Venice (2007) (here I have a bias) "A century of Cosmology: Past, Present and Future", to which many[c], but not all, of the leading scientists working in the field participated.

I added a few details to the talk given in Stockholm since I presented the same talk to students and the few additions may ease their understanding. Furthermore and under the suggestion by Public Outreach and Education I described now and then the way I felt.

I. **Introduction.**

---

[b] Various good reviews have been published, among these that by H.J. Rood (1981) Rep. Prog. Phys. 44 1077.
[c] Prof. Avishai Dekel, for instance, did not participate because of a misunderstanding due to the organization and a later conflict he had with other commitments. His work had been fundamental also in the early days for the understanding of the large scale peculiar motions triggered by the large-scale distribution of mass. Catherine Cesarski, Massimo Tarenghi, Adam Riess, Saul Perlmutter and a few others were unable to submit the proceedings paper.

Based on photographic studies, that is on a two dimension distribution, the way we conceived the distribution of galaxies in the early seventies was essentially due to three great pioneers of extragalactic astronomy: Harlow Shapley, Edwin Hubble and Fritz Zwicky. Shapley showed that distribution of galaxies was very irregular and we could observe large aggregate; clumps and Superclusters[d] of galaxies. This is seen, for instance, in the plots of the Shapley – Ames Catalogue, Figure 1.

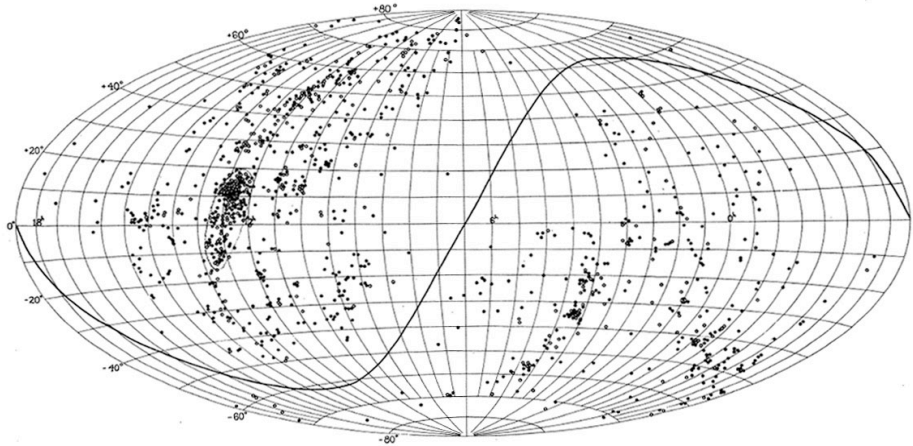

Figure 1. The distribution of galaxies observed by Harlow Shapley and Adelaide Ames (1932)[4]. The plot shows 291 nebulae brighter than 12$^{th}$ photographic magnitude and 734 between 12$^{th}$ and 13$^{th}$ magnitude. The Virgo Cluster and the indication of the Local Supercluster are clearly visible.

The problem with the large surveys carried out by Shapley on the rest of the sky was that the magnitudes, due to the limited facilities of the time, were rather inaccurate. Quite often when we, later on, tried to use the catalogue of galaxies made by Shapley to observe the Southern sky (the region of the Horologium in particular) we could hardly use the published coordinates to identify the galaxies and some were simply not there. On the other hand his conception of the Universe was visionary. Hubble was, and perhaps is, the most influent and known astronomer of the time (very capable also in public relations) so that some of his views were easily accepted and conditioned the thinking for a long time. Hubble believed that all galaxies, groups and clusters, are randomly distributed on the sky [a concept that after all was quite in agreement with the believe of the time and with a uniform distribution of matter reflecting, to some extent, the concept of the ancient Greeks of a perfect and unchanging Universe]; Shapley agreed, in a sense, with this view over great volumes of space emphasizing however the presence of clumps of about 70 – 100 Mpc. For the first time, even if not discussed at length, is introduced the concept of a characteristic scale length.

---

[d] Shapley talk also about the Metagalaxy that comprehend a rather large part of the Universe while the inner Metagalaxy refers more or less to the Local Supercluster (Shapley states 70 – 80 million light years for this local part of the Universe).

Zwicky on the other hand, one of the most innovative observational astronomers of the last century, conceived the distribution of galaxies as a uniform sea on which were superimposed large clusters of galaxies. His view (1972) is illustrated in the sketch reproduced in Figure 2[e] where it is clearly shown that on top of a uniform distribution of field galaxies we have superimposed the clusters of galaxies. Indeed the idea of the existence of a field was very strong and pervaded all thinkers of the time.

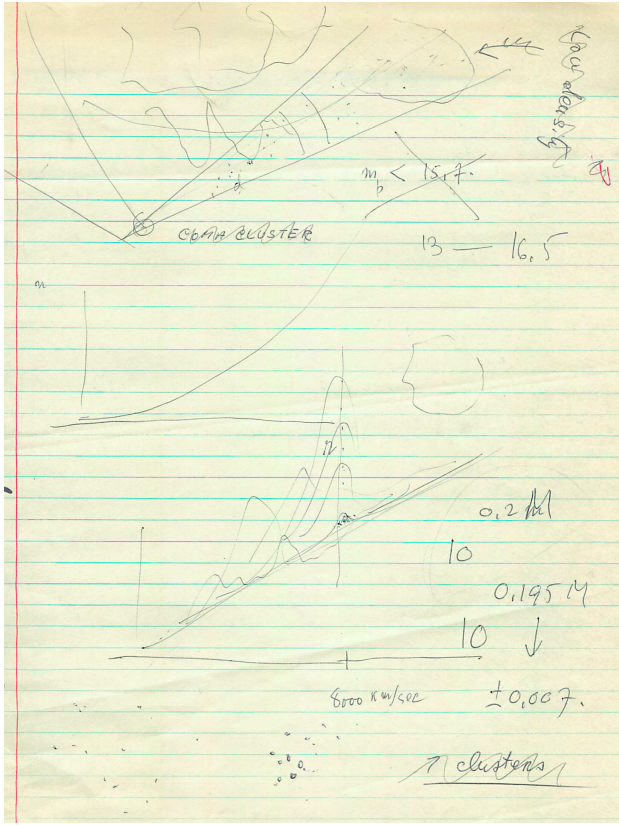

*Figure 2. A sketch made by Zwicky in Pasadena (Caltech) during a discussion I had on the distribution of clusters of galaxies. The top plot shows the cone diagram from the observer and the bottom plot the distribution of field galaxies on which clusters are superimposed.*

Oort (1981)[5] in his article "Some notes on my life as an astronomer" writes: "Curiously, the relations with de Sitter never touched the subject of the Universe, in which I became so strongly interested in later years. The Einstein-de Sitter Universe and all that pertained to the subject was hibernating at that time". On the other hand as probably the most famous astronomer of the past century he was invited at the Solvay Conference (1958)[6] to give a talk on the Large Scale Structure of the Universe where he pointed out the inhomogeneous distribution of galaxies.

He begins writing: " One of the most striking aspects of the Universe is its in-homogeneity" and later " … there is no such a thing as a regular <<field>> on which the structures we see are superimposed". Of course he is guided by the distribution of

---

[e] In 1972 I visited for about two weeks Prof. F. Zwicky in California to discuss compact galaxies and clusters of galaxies. It has been a full immersion; I was talking with him most of the time (I really was mostly listening) including during the exciting lunches at the cafeteria of the Institute.

I came back from that visit, that I also discussed with Herb Rood, with the firm persuasion that we should find ways to answer questions that were disturbing since some time: how large are clusters of galaxies (Herb was deeply involved in the field and fascinated by the dynamics of clusters) and how uniform is the distribution of galaxies [all standard cosmologies were based on a uniform distribution].

galaxies depicted in the shallow catalogue (1932 - 1938) by Shapley and Ames (local and only bright galaxies and yet indicative) and misleading however when dealing with the concept of homogeneity and its related scale length. In this presentation he states, page 167, the Hubble constant (the cause of a big controversy at the time and for some time to come) in the Virgo cluster is the same as in the rest of the Universe (the observations were not accurate enough and the fluctuations in the Hubble flow caused by the perturbations in the matter density distribution were not yet known). Peebles and his collaborators developed the sophisticated autocorrelation theory to track down the distribution of light in the Universe and Avishai Dekel developed the theory and analysis [the potent program] to map the distribution of matter. Oort refers also to the monumental work by Shane and Wirtanen[f, 7] and to the analysis that was carried out by Neyman, Scott and Shane[8]. The analysis evidenced a likely characteristics scale of four – five degrees. It is also remarkable, in spite of the little and local evidence, of the time, how close Oort was to the picture of the World we understood much later. However not based on observations derived from a fair sample of the Universe or statistically significant. His attachment to this topics is shown by the article he wrote in occasion of the inauguration of the ESO headquarter in Garching (May 1981)[9] and by his opening talk at the IAU Symposium N.104 held in Crete in 1982 (see however in the same volume Chincarini et al.[10], and Thompson and Gregory[11] and references therein, for a more accurate history of the detection and definition of superclusters). In 1981 he wrote an innovative paper testing the clustering on large scales of quasars.

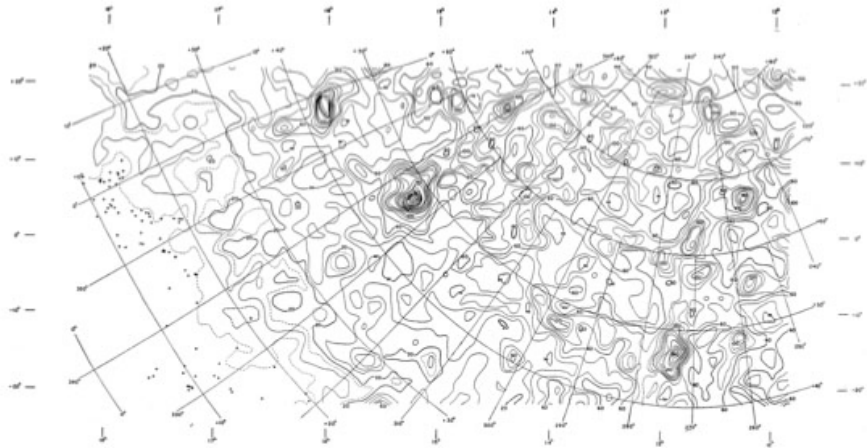

*Figure 3. Equal density contours of one of the regions observed by Shane and Wirtanen (1954). Counts are smoothed over 1 square degree.*

Gerard de Vaucouleurs in 1958[12] using redshifts and photometric data available at the time proposed the concept of a local Supercluster. The problem with that model is that it is a dynamical model with a distribution of mass (galaxies) in a state of differential rotation and differential expansion[g]. This cosmic unit is too big (about 30 Mpc always according to De Vaucouleurs) to reach equilibrium in a Hubble time. I always was convinced that De Vaucouleurs kept the view of the Local Supercluster as a dynamical unit (see De Vaucouleurs 1976[13] and 1970[14] where the mass of the differentially expanding and rotating super system is derived using the rotation of the supercluster) even if he was aware of the complexities of large scale clustering[h].

---

[f] The Lick Observatory Carnegie double refractor was designed, and built, for a proper motion project conducted by Vasilevskis. Shane had the fantastic intuition that the telescope could be used for surveying the distribution of galaxies. This pioneering work formed a fundamental database for years to come.

This drives us to pay attention to his paper on clustering written in 1971[15]. Here he refers to the work by George Abell who used cells of various sizes to estimate the clustering properties of clusters of galaxies. Large scale clustering "fluctuations are still indicated on a scale of $\Lambda \sim 100$ Mpc" is being recognized referring in particular to the correlation analysis by Kiang and Saslaw (1969)[16]. In this analysis some power is still detected at correlation lengths of about 200 Mpc and more. Yu and Peebles (1969)[17] tried to detect evidence of super-clustering from the analysis of the distribution of clusters of galaxies. Their result is: "Our results suggest that any tendency toward super-clustering is scarcely above that to be expected for a random distribution. On the other hand, our results cannot rule out the existence of some super-clusters. …". Later on Neta Bahcall and Raymond Soneira (1983)[18] measured the autocorrelation function to a rather high degree of robustness evidencing indeed not only clustering but also measuring a correlation scale length five time larger than that estimated for galaxies.

Back to the paper by Gerard he also refers to the old work on clustering and related conception (not always supported by reasonable data) of higher order clustering: III, IV etc. That he was convinced of this type of hierarchical clustering it was evident in the various discussions we had and in particular a phone call (about 1974) we had while at the University of Oklahoma I was working, in collaboration with Herb Rood, on the redshift surveys in the regions of the Coma Cluster. His view was supported also by the very interesting correlation he found between the size of a structure and the density, Figure 4. This plot is, to some extent, very close to what we would expect from the fractal distribution of objects as developed by Mandelbrot. Later various scientists defended the fractal distribution on all scale lengths even if that would lead to extremely large low-density structures and to extremely low density Universe. Observations did not support this view however. The simple fact is, after Rood and I realized the existence of a filamentary structure and voids[i], that there is no evidence of a higher order clustering and the mean density of the Universe is that measured over enough large volume of the Universe. This concept was in full agreement with the work by Peebles and his collaborators.

---

[g] However during the Tallin meeting at the question by Zeldovich: "Is it due to rotation? " [He was referring to a flattening of 0.4 – 0.5 mentioned by De Vaucouleurs] the answer was: "No, no in the sense of a Newtonian spheroid in centrifugal equilibrium"
[h] On this topic we had various discussions since he had an excellent knowledge of the literature. We had some disagreement on the Local Supercluster and higher order clustering however.
[i] Gregory and Thompson after they started to observe to increase the redshifts in the Coma A1367 region came to similar conclusion. As I specify later we were often exchanging ideas and data in the endeavor Rood and I started.

Zwicky always confuted the view of second order clustering, and to some extent that of super-clustering because of a different concept he had about it. His concept of super-clusters was not that of an over-density but that of a clusters of clusters and so on. He was used to state that clusters of clusters of galaxies of the Coma – type, that is a cluster of clusters where the member clusters act as the galaxies in a cluster of galaxies, do not exist. He was obviously right, but only according of his definition of super-cluster, an extremely large dynamical unit. While he would openly discuss this point with many astronomers, in a remark he made after the talk De Vaucouleurs gave in Padua[j] in 1964[19] he states: " I was glad to hear that the nonexistence of genuine clusters of clusters of galaxies, which I have arrived at from the analysis of the spatial distribution of about 10000 clusters of galaxies, is now being admitted by the speaker".

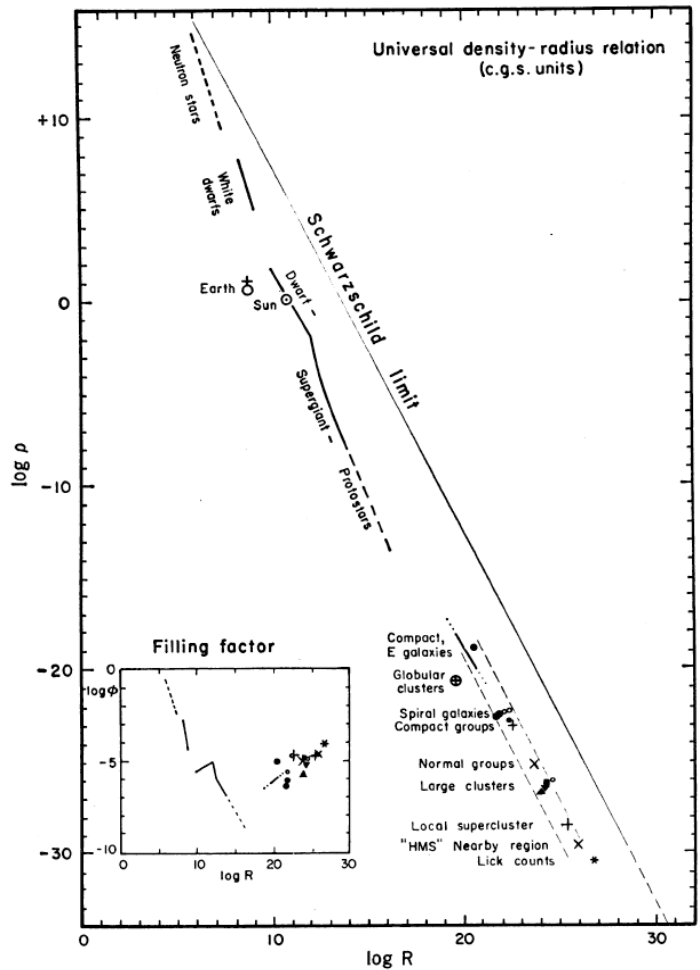

*Figure 4. The density – radius relation give the maximum average density of matter (grams per cubic centimeter) in a spherical volume of radius R (cm). The inset shows the filling factor and the continuous line the Schwarzschild limit. The concept he develops is that clusters of galaxies occur on all possible scales with no preferred sizes.*

Perhaps it is only semantic and the difference, as I said, is in the nomenclature.
What he calls super-clusters are simply large medium compact or open clusters of galaxies within which there exist a number of more or less pronounced condensations of galaxies. Etc." For the conception Zwicky had of the distribution of galaxies I refer again to Figure 2 and only in part the different of opinions were due to semantic as remarked also by De Vaucouleurs (1970)[14] in an excellent review article. This article, with which I conclude the description of the state of the art in the early seventies, is of great interest under various aspects and reflects the state of confusion on the large-scale distribution of galaxies at the time. Indeed accurate data were lacking and the old ideas, due to photographic surveys and sometimes to a preconceived cosmology pervaded the approach to Nature. Part of what we did not know was supplied by believes and kind of a metaphysics, metagalaxies, hypergalaxis etc. Indeed Gerard concludes his article with the following statement: *"It seems safe to conclude that a unique solution of the cosmological problem (he refers here to the large scale structure) may still elude us for quite some time!"*

---

[j] This was a meeting in occasion of the fourth centenary of the birth of Galileo – Galilei (1564 – 1964) and the proceedings were published in 1966 by G. Barbera Editore Firenze.

**The point I want to make with the above discussion is that the presence of a uniform field of galaxies on which clusters were superimposed, Figure 2, was at the time the model accepted and to some extent demonstrated by the counts of galaxies done in various parts of the sky. Agglomerates and clumps were perturbation superimposed to this field. I'll refer later to important theoretical work[k].**

II. The readiness for spectroscopic surveys.

*The challenge is not the formation of new ideas
but rather the obliteration of the old ones that pervade our minds
Adapted from John Maynard Keynes (1935)*

A milestone of extragalactic astronomy and cosmology is the paper by Humason, Mayall and Sandage (1956)[20]. Their catalogue consists of 620 nebulae (galaxies) observed (redshifts) at Mount Wilson and Palomar and 300 nebulae observed at Lick Observatory Mt. Hamilton.

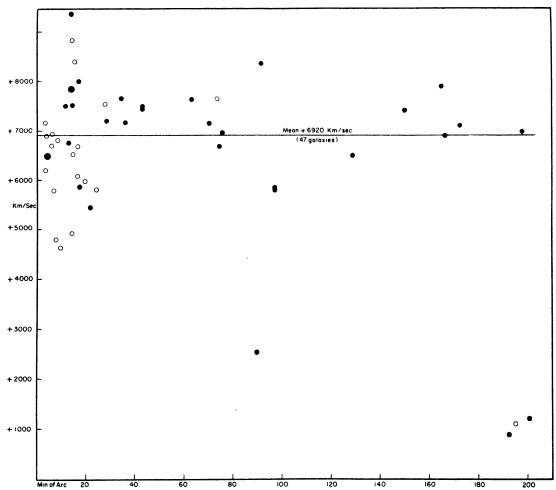

*Figure 5. Open circles indicate Mt Wilson - Palomar observations, filled circles Lick Observations (from Mayall 1960)[21]. This characteristic plot will remain very similar even after a large amount of data will be added.*

To have some flavor of the state of the art at the time: the spectra were obtained during the 20 years interval from 1935 to 1955, exposure times are in units of hours (even if the galaxies were very bright according to modern standard) up to maxima of about 14 – 17 hours and the Hubble constant the authors

---

[k] And for a theoretical study on field galaxies see "Field galaxies: Luminosity, redshift and abundance of types. Part I. Theory" by Neyman and Scott (1961)[22].

derived at the time was H = 180 km/s/Mpc.

In the sixties spectra of galaxies were still hard to get due to the low sensitivity of photographic plates. To make up for the low sensitivity of the detectors astronomers designed very fast spectrographic cameras and we were obviously working at very low dispersion. However something was coming up. N. U. Mayall realized the great potentiality of the electronic photography as developed by Lallemand for getting spectra of faint objects. In 1960 Mayall wrote: " Among the fields of observational astronomy, that of extragalactic spectroscopy is one that stands to gain much from the application of electronic photography, as developed at the Paris Observatory by Prof. Lallemand. He has shown that sensitivity gains of 50 to 100 are possible. This means that exposures previously requiring hours may now be obtained in as many minutes. "

Merle F. Walker[l], one of the very best observers I ever met, imported this equipment in the US at Lick Observatory Mount Hamilton, since he fully understood, in spite of the difficulties and problems in using it, that the painstaking effort would be compensated by the unique results and discoveries. The instrument would allow taking either spectra of faint objects with reasonably good spectral resolution or spectra of highly variable objects with high temporal resolution. Moreover, gains of this order permit the use of higher dispersion. Indeed the results obtained by Merle F. Walker were by far ahead of time! Meanwhile RCA and Westinghouse[m] were developing much user friendlier images tubes electromagnetically and electrostatic focused (the Westinghouse tube used also fiber optics output, a technique that would allow multistage devices)[n]. The Carnegie Image Tube - RCA became a standard tube for the astronomical community and was used at Kitt Peak National Observatory and in other facilities (Kent Ford, in collaboration with Vera Rubin, at the Naval Observatory for instance). The experience I gained and developed at Lick Observatory came at hand at the Mc Donald Observatory (University of Texas) where I could experiment and observe with some of these, including a three stages ITT tube.

---

[l] In 1964 I went to work with Merle at Lick Observatory (University f California) as a Post Doc to assist him with the observations and the development of the electrographic Lallemand's camera. I became one of the few experts in the field. Later at the Johnson Space Flight Center (1969), NASA, I developed a laboratory using the electronic camera made by Jerry Kron (I had a observing run at Mauna Kea with it) and finally at McDonald Observatory I developed, in addition to continuing observations with the Kron camera, industry manufactured image tubes and related spectrographs. This know how gave me expertise that opened the way to the observations of Clusters of Galaxies in the group of Thornton Page (1969) and shortly after the fortunate collaboration with H.J. Rood.

[m] While designing a image tube intensifier for space I visited with Tom Giuli, as a NASA affiliated, these industries and in addition to various developments in the field we got to know they were developing some new detectors kept, however, very secret; likely they were developing the CCD.

[n] I limit myself to mentioning the main stream of devices used for light amplification, however various Institutes in the US and UK developed other devices that gave in various applications excellent results. Among these the Vidicon as developed at Princeton, the Wampler's scanner used at Mt Hamilton and later in Australia and many others.

Herbert J. Rood[o], then at the Wesleyan University, got his PhD at the University of Michigan with a thesis on the Dynamics of the Coma Clusters of Galaxies. Herb was fascinated by cluster dynamics and by the complexity, or simplicity, of such large cosmic units. Rood, Page, Kintner and King (1972)[23] in a very detailed study of the cluster (I know it took quite sometime of interaction among the authors so that the results largely preceded the publication) conclude that the missing mass is present but is not in the galaxies and that the distribution of the missing mass had to be rather similar to the distribution of visible galaxies. This result, in spite of the fact that Zwicky was talking about Dark Matter[p] [missing mass and dark matter here are used in purpose in different context to stress the difference between the two concepts] already in 1933 is fundamental since at the time the idea of stability and missing mass[q] was not fully accepted by the community. To deal with these problems in 1969 – 1970 we started to measure the redshift of cluster galaxies, and the region of Coma in particular (Coma was considered the prototype of relaxed clusters) to accurately determine the M/L and other cluster (and galaxies) parameters. But then, as I mentioned earlier, as a second step the two main questions we wanted to eventually answer were: a) how large is a cluster[r] and b) how homogeneous (on which scale) is the Universe. These goals were strengthen after visiting Zwicky in Pasadena.

**III. The early spectroscopic work**

Herb and I started to collaborate in 1969 and published in 1972[24] the first catalogue of redshfts based on spectrograms obtained primarily in 1970 with the Carnegie image tube Cassegrain spectrograph attached to the KPNO 84-inch telescope. That first survey is heavily weighted on spectra about the Perseus clusters since I noticed that only three redshifts were known in that region (Humason, Mayall and Sandage catalogue, 1956).
Rood and I published one of the main results of our work in an article Nature in 1975[25], but see also Chincarini and Rood 1976[s,27], where the first evidence[t] ever is given about the existence of super-clustering using redshift as the third dimension. At the time, in spite of the plot made by Zwicky, Figure 2, that showed the distribution of galaxies using a cone diagram (solid angle) as viewed from the observer, we preferred at first to go with Cartesian coordinate for simplicity. Indeed we did not think much about it.

---

[o] I joined Thornton Page at the Johnson Space Flight Center, Houston, in 1969 following a telegram asking me the following question: "Do you want to come here and put a telescope on the Moon or stay in Germany observing clouds". At the time he and Rood had started a program at KPNO on measuring velocities of galaxies in the coma cluster. In one of the early run at KPNO I substituted Thornton (he was too busy planning his telescope for the Moon) and from then on Rood and I worked together.

[p] J.P. Peebles in the book "Physical Cosmology" by Princeton University Press (1971) has an interesting discussion on the mass in the Universe and this reflects in part the state of the art at the time (summary page 115).

[q] Discussions at the University of Texas Austin at the time David Schramm and Beatrix Tinsley were there would often refer to a concept of missing light as well.

[r] On the size of Clusters, and in particular on the size of the Coma Cluster, there were contradictory statements between astronomers. Zwicky, who used also the 18 inch Schmidt telescope for his surveys insisted on a very large size, see his Morphological Astronomy, 1957, Edited by Springer Verlag.

[s] Herb was inspired by Yahil's paper, however, thanks also to the referee, we (Herb was the final writer) modified part of the discussion.

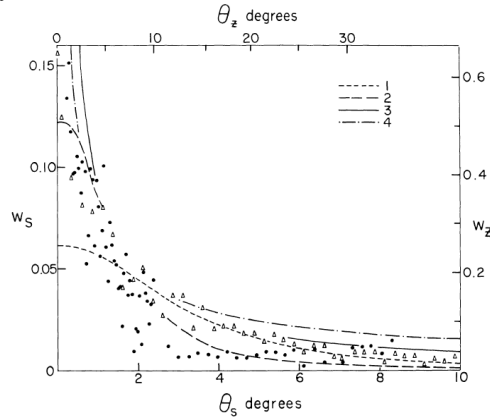

*Figure 6. The angular cross correlation function for the Zwicky and for the Shane-Wirtanen Catalogues (Peebles and Hauser 1974)[28]. Both the coherence length and the exponent of the autocorrelation function are very close to the values measured today on larger samples. In the analysis Peebles and Hauser use the Luminosity Function derived by George Abell (1962)[29].*

Rood got to know, and mentioned to me during a working visit, about the work by Jim Peebles[u] based on the auto-correlation function analysis. We contacted him and mailed the work we were doing at the time. Jim was very encouraging about our work and that stimulated and excited us even more. Following the analysis he and his group made on the catalogue by Shane and Wirtanen, Figure 6, and on the Jagellonian Catalogue (1975)[30] – see Appendix I - I started to study the theory of the autocorrelation function analysis, with the high respect a rough observer has for the theory[v]. The point that most stroke me at the time was the fact that in the autocorrelation it was implicit the lack of a center or of a reference point so that this type of statistical analysis reflected very clearly one of the fundamental characteristics of the Universe: lack of any center or reference point, a way to address the concept to students. On the other hand we also realized, even if in a foggy way (I mean not carefully justified by us theoretically), that the two points correlation function does not say much about the characteristics of the patterns of the galaxy distribution and even higher order correlations, I learned later on, have difficulties in giving details on the structures. This gave us more confidence that the simple observational work, that required a lot of observing time anyway, was very valuable and fundamental to the understanding of the distribution of galaxies. Rood and I, however, had to pass through the Time Allocation Committee and compete for time not easy to get. Indeed it is also because of these limitations[w] that we started to have soon after the first results some problems in getting the time we needed and I decided to look for other ways to observe[x]. I started the extragalactic LSS work at the radio telescope of Arecibo, about 1975.

---

[t] While searching the web for the images related to the distribution of galaxies according to the catalogue of Shane and Wirtanen I came across the article (Science, Vol. 219, 1050, 1983)[26] by M. Mitchell Waldrop assigning to Einasto (to my knowledge his group never carried out a redshift survey) the pioneer work on Large Scale Structure in 1970 coupling it, at least the reader gets this impression, to Peeble's comments on the value of redshift surveys! It is not clear where from Mr. Waldrop got this. From the statement by Peebles he quotes (but see footnote 13) I get the impression he confused work and scientists or simply did not know how things developed".

[u] I was so impressed by Jim Peebles and his work, that since then I considered him one of my "reference point" in cosmology and often I asked his advise and opinion. In the paper "Cosmic Virial Theorem" Ap.J. 205, L109, 1976[31] he stated: "For a more believable discussion of Figure *1 [his Figure 1]* we need *(<v²(r)>)* based on a "fair sample" of galaxies. Redshift measurements in selected areas of all galaxies at *m < 15* (Zwicky *et al.* 1961-1968)[32] probably would be adequate because the correlation functions derived from this catalog agree with the results from deeper surveys. Of course, the redshift data would have other important uses, as has been vividly demonstrated by Chincarini and Rood (1975). The project is formidable but possible if it were agreed that it is worth doing".

[v] I asked Jim Peebles to let me know when his book "The large scale structure of the Universe" would be published and I bought it right away. I later spent a summer at ESO writing a Fortran program to compute the two point $\xi(\sigma,\pi)$ correlation function.

From this early experience, thinking about it, I learned it is extremely important to be in close contact with good theoreticians or to know theory in all details as well. Rood, and especially I, was missing during the early work such opportunity.

Bill Tifft was interested in redshift observations for studying and interpreting some correlations he detected in the Coma cluster between magnitudes and redshift (1972)[33]. Bill interest in spectroscopic observations started the work of Gregory and Thompson who got busy in the study of the distribution of galaxies[y]. Tifft, Gregory and Thompson were in excellent contact with us (we all used telescopes in the area of Tucson Arizona) exchanging information and data so that in my opinion our ideas and concepts in those years grew partly together and all together gave what I believe was a fundamental contribution to the field. See for instance their 1978 paper[34].

Another fascinating thematic was debated at the time; I refer to the work and ideas of Chip Arp[35] and Geoffrey Burbidge, among others, who were supporting the non-cosmological redshift interpretation[z]. Two test cases were the groups Stephan quintet and the Seyfert Sextet, the first with a discrepant redshift lower that that of the other galaxies and the second with a much higher redshift. This was proving, according to some, that something wear was going on. Rogers Lynds (1972)[36] obtained on the other hand the redshift of galaxies located in a small region of the Stephan quintet showing that quite likely we were dealing with the superposition of groups. Chincarini and Rood (1972 – A.J. 77, 4)[24] obtained a similar result for the Seyfert sextet, Figure 7, in their small catalogue. Donald Martins and I decided, therefore, to look deeper into this problem in view that I already had an observing program on that region of the sky.

---

[w] One of my proposal at KPNO to survey Superclustering was rejected (this obviously we know could happen). Independently of such inconvenient, and soon after it, Tom Kinman (Tom was a fantastic person and I knew him since the time we all were at Lick Observatory on the top of Mt Hamilton) wrote to me asking to send some material about the work Rood and I were carrying out since KPNO wanted to present it as a highlight to the NSF. I answered (copy to the Director, Geoffrey Burbidge if I recall properly) in an upset mood evidencing the contradiction with the TAC negative evaluation and finally did not send anything! I realized only afterwards my stupidity; stupidity that literally fits to the description given in the book by Carlo M. Cipolla "Allegro ma non troppo con Le leggi fondamentali della stupidità umana" (Allegro ma non troppo with the fundamental laws of the human stupidity).

[x] In this period I was also full of personal problems that triggered changes both in life and in science.

[y] Bill was already interested in observing Coma to support his interpretation about what he called the redshift – magnitude pattern. Herb and I were observing at KPNO and now and then we would meet with Bill's group. I recall Herb, he always was aimed to get as many observations as possible in the region of the Coma Cluster and other regions, would suggest to push survey work given the availability at Steward Observatory of the new Bok 90 inch telescope (1969).

Indeed I had the opportunity of using telescope commissioning observing time because I rebuilt a spectrograph at McDonald Observatory for the Cassegrain focus of the 107 inch telescope and I could use it for science (the redshift surveys I started with Herb Rood) some of the test time. Astonishingly, while observing and developing the photographic plates[aa] I noticed that the observed galaxies had redshifts grouped in rather well defined ranges (by eye and knowing well the comparison and sky spectrum it is possible to estimate the redshift with the accuracy of a few hundreds km/s).

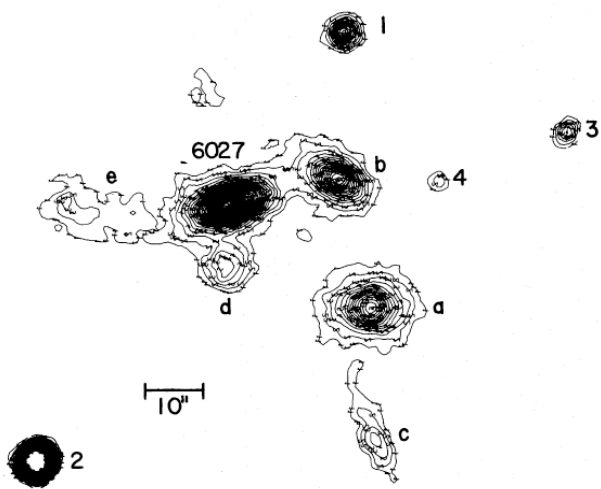

*Figure 7. The photometry by Martins and Chincarini (1976)[37] made, using the electrographic camera developed by Jerry Kron at Mc Donald Observatory excluded the interaction of the discrepant redshift galaxy (d) with NGC 6027. The velocities are (km/s): d = 19809, 6027 = 4446, b= 4147, a=4292, c = 4503. Spectra were also obtained in the whole surrounding region.*

Chincarini and Martins[bb] (1975)[38], triggered by these observations (see the paper for details) checked also other samples of galaxies for which we had redshifts. We discovered that the distribution of redshifts, over completely different regions of the sky, is segregated in groups. This "segregation in redshifts", as we called it in that paper, was again indication of super-clustering and, I realized it soon afterwards as a natural consequence, holes or voids[cc].

---

[z] A simple and fast reading on this matter is: The redshift Controversy 1973, edited by G.B. Field, H. Arp and J.N. Bahcall, publisher W.A. Benjamin Inc.
[aa] I normally developed each plate during the night after getting a set of spectra on it.
[bb] We first submitted the paper to the Astrophysical Journal Letters because I considered it an interesting result to communicate quickly, however Gerard De Vaucouleurs was the referee and he came to my office (at the time we were both in Austin and on the same floor of the Astronomy Department) and explained the article should go to the main journal rather than to the ApJ letter as we had done. I obviously was interested to publish it soon and I accepted the wasting of time (I could not do anything anyway) transferring it to the ApJ main journal to avoid any further delay.

It happened that in those years Franco Pacini organized a meeting in Italy at the Accademia dei Lincei[dd]. I presented (May 21$^{st}$ 1976) the evidence of the superclustering in the region of the Coma & Hercules Clusters, Figure 8, that we obtained from the redshift distributions (Franco never published the proceedings even if they were planned, lack of time I guess). Martin Rees, looking at the plots and histograms of the redshift distribution, asked (the visual impression was clearly showing the large scale aggregations and the empty regions I was talking about), whether there could be a bias due to the magnitude-limited sample. Back home I kept thinking on the problem to see whether I could give a simple statistical evidence of the distribution. I came up with the Nature paper "The clumpy structure of the Universe and general field" (1978)[ee, 1]. **This is the first statistical evidence of the Voids.** Later we tried to do a more complete statistical work, Vettolani et al., 1985[39], but I felt we did not find the best statistical tools for that task.

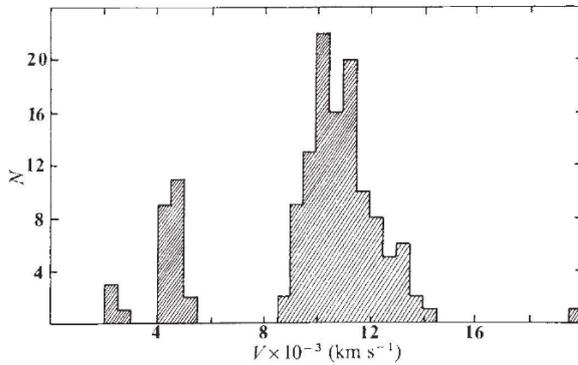

*Figure 8. Histogram of the velocity distribution for galaxies in the region of Hercules Supercluster. We had observed this segregation of redshift also in Coma (Chincarini & Martins 1975[38], Chincarini & Rood 1975[25] Nature) on the other hand the data we were collecting on Hercules seemed a better case for the computation (sample ~ complete to 15.7) and the observed gap was statistically significant and robust.*

The number of galaxies expected in a defined velocity range [at the time we used the Abell's Luminosity Function] can be computed

$$N(V_1, V_2) = \Delta\Omega \int_{V_1/H}^{V_2/H} x^2 \phi(x)\, dx$$

after normalizing the distribution in velocities using counts of galaxies complete to a limiting magnitude. I used the normalization:

$$\frac{N(0, \infty)}{\Delta\Omega} = 1.905\, A\, D^* = C^{0.6\, m_{\lim}}$$

---

[cc] Rood and I used the word "Voids" in the context of the distribution of galaxies obviously after my Nature letter (1978, submitted in 1977) – Earlier I called them "holes".

[dd] May 20 – 22 1976

[ee] Herb Rood decided not the coauthor the paper, I discussed it with him, because a) He thought I should do it by myself and b) with our work and findings we were so used to the reality of visual inspection that probably he thought the finding was obvious.

C is estimated using counts in the Zwicky catalogue, $D^* \equiv 10^{0.2\left(m_{lim}-M^*\right)}$ with $M^*$ from the Abell's Luminosity Function and A the needed normalization of the distribution in velocities.

Meetings and proceedings, as any other script, are quite often biased documentation of the evolution of science since they do not reflect the communications among scientists and not all scientists, who are relevant to the evolution of a subject, participate to them. Nevertheless they tell us something so that I will refer to some of those I believe relevant to this topic. To some extent they reflect the evolution of ideas.

At the time I considered the meeting held in Tallin (1977) important because I could illustrate the results Herb and I were achieving. Early enough I submitted an abstract to illustrate superclusters and in particular the "Hercules" Supercluster[ff]. In a retrospect I consider that meeting important also because it shows to some extent the state of the art and the confusion some had on the matter. It also stimulated further work by others (see Chincarini The Messenger 1981)[40].

Einasto and his group before Tallin were mainly interested in what they called "hypergalaxies". A very foggy concept even if it may be close to what we knew, for instance, about the Milky Way and satellites albeit the dynamics and the geometry (see for instance Einasto 1978[41] and the questions and answers therein). In any case after 1976, and more following the Tallin meeting, the group headed by Einasto started to pay attention to Superclusters using the data that were becoming available. In Joveer et al. (1978[42] Tallin Proceedings but see also MNRAS 1978, 185, 357) the samples used are not complete, however, in a statistical sense and some of the concepts, known from the literature and meetings, are not supported by robust data.

John Huchra at this meeting mentions the beginning of the large survey (at the time yet too shallow with m < 13.0 so that at this meeting he does not discuss Superclusters) planned by Davis, Geller, Tonry and himself. Tonry and Davis in the first paper of the series describe the data analysis while the first part of the survey, Figure 9, is discussed in Davis et al., 1982. This is the beginning of the very large surveys to study the Large Scale Distribution of galaxies and John played a fundamental role in the field. In 1982 George Abell[gg] and I organized a meeting in Crete, IAU Symposium N. 104[hh]. In the

---

[ff] I was particularly excited because the progress we had made on the topics and the congratulation we had verbally from Zeldovich. Unfortunately, my slides (home made) were terrible (as I realized and was told also by Cesare Perola). Likely Bill Tifft and his group enjoyed as well Zeldovich's compliments.

[gg] His contribution to the field has been fundamental under various aspects. He had planned the extension of the cluster catalogue to the South but unfortunately he died before starting the work. In his honor and in agreement with H. G. Corwin and R. P. Ollowin (a student of mine) I decided to write a NSF proposal and carry out the work on his behalf. I did not co-author the catalogue since Harold and Ronald following George's ideas did all the work.

[hh] During the organization of the meeting I asked advise to Martin Rees about neutrinos since at the time many were considering neutrinos could be the "unseen" mass. He told me that was probably the peak of the popularity for neutrinos and things would likely

same years I also lectured at the III Brazilian School of Cosmology and Gravitation organized in Rio de Janeiro.[ii] In Crete it appeared clear that the concept we had in mind about clustering in the seventies was what was confirmed by larger surveys, CfA in particular. The AAT at the time just begun and could not yet describe the filamentary structure interconnecting clusters as implicit in the question by Richard Ellis[jj] after my presentation:

*"Would you comment on the existence of large structures as general features of the galaxy distribution? Redshift surveys in "interesting areas" may reveal such structures, but their significance as general features can be assessed only by performing deep surveys in randomly chosen directions.* ***The AAT survey (Bean et al., this symposium) do not show statistically convincing evidence for these large-scale features".*** The concept of randomly chosen region of the sky or all sky survey is, of course, correct; on the other hand the fact that the same features were observed in magnitude limited samples selected in different regions of the sky, see the answer, was a robust evidence of the findings.

***Going back to those years and quickly scanning the work I feel, as I did feel at the time, that the Nature Letter (1978) remains a milestone, a change of view. It marks the new vision of the distribution of matter and the beginning of new redshift survey approach to the problem followed by the unique and powerful CfA survey that quickly gathered copious and fundamental data to detail the Cosmo.***

---

change and decline in the future. He was right. I also wrote with some excitement (he still was one of my Gods of the time I was a student) to Prof. Bruno Pontecorvo asking him to tell us about his work. He could not come because he was ill but, with the modesty of the great men, he appreciated the invitation and the fact that we highly considered his work.

[ii] http://www.cbpf.br/~cosmogra/Schools/IIISchool.html, I never received the proceedings and however I mailed to many scientists working in the field the article as printed at the University of Oklahoma (I have now a digitized copy). The listing does not include all the scientists who lectured since I recall among others Vittorio Canuto, E.M.Lifshitz (?) and other well-known scientists were lecturing. In those proceedings I also mention the obvious fact that we expected a component of motion, from the CMB data, toward the Hydra – Centaurus Supercluster.

[jj] See the discussion page 165 of the proceedings edited by Abell & Chincarini (1983). Richard Ellis is one of the main contributors to the establishment of the observational cosmology in the UK. This represented, in my opinion, the rise of excellent observational expertise, following what probably was one of the very best theoretical centers of cosmology, and an important gap in the UK astronomy was filled.

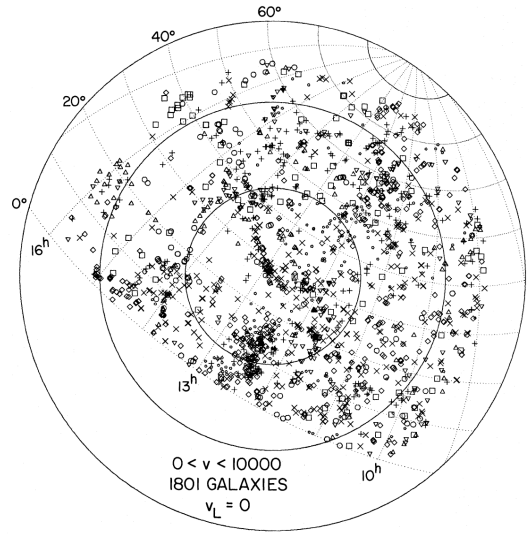

*Figure 9. The first complete sample of the CfA survey consists of 2400 galaxies brighter that photographic magnitude 14.5 [at the timewe all used the catalogue of galaxies by Zwicky. In this early plots it is hard to clearly see voids, but some structure is visible. This very dedicated survey became the reference point in the coming years and proceeded very well thanks to the dedicated instrumentation available. Davis et al. (1982)[59]*

Previous attempts were great and based on highly sophisticated theoretical work, however could not get to the point. To name a few between the late sixties and early seventies we should refer to the work by Nyman, Scott and Shane (1956)[43] who applied the statistical analysis to the catalogue of Shane and Virtanen, the work by George Bell on the analysis of clusters and the statistical analysis by Kiang (1967)[44] and Kiang and Saslaw (1969)[16] on the clustering of clusters of galaxies[kk]. These papers are milestones in the development of our understanding.

Fall et al. (1977)[45] and Soneira and Peebles (1977)[46] investigating the characteristics of field of galaxies at the end of their papers refer to the needs of redshift surveys. We, Rood and I, started in 1970 working on Clusters and from there the curiosity, and some discussion with Zwicky, arrived to a new picture on the distribution of matter before the end of the seventies and this was a good complement to the fundamental work by Peeble's group.

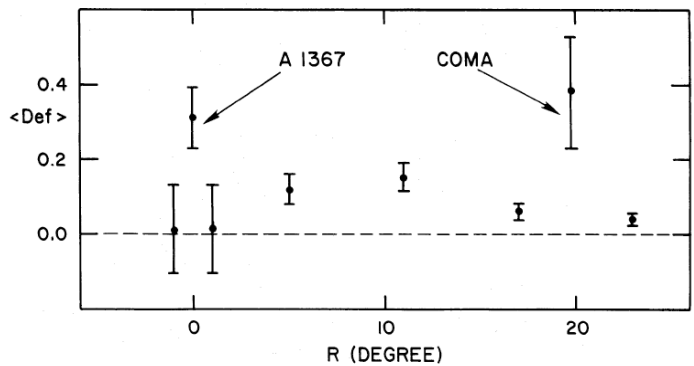

*Figure 10. The first results we obtained at Arecibo on the deficiency of galaxies in the clusters Coma and A1367 (Chincarini, Giovanelli & Haynes 1983)[47]*

In the coming years structures and details on the distribution became progressively clearer (see also Chincarini and Rood, Sky and Telescope 1980) as discussed in the 80's (IAU Symposium 124 in Benjing) where the CfA survey was already playing the big role and later with the many fantastic surveys that built the rich archives and simulations we have today, among these of primary importance the SDSS that led to an extremely large amount of new results and understanding. The basic of

---

[kk] I refer the reader to the proceedings of the III Berkeley Symposium for an enlighten reading.

progress and the concern of those planning surveys have always been the completeness and size of the sample, volume or magnitude limited as Rood and I started, that would enable to do statistics in a fair sample of the Universe.

I refer also the excellent work by Gregory and Thompson (after 1975 we kept in contact and, especially Herb, exchanged data and information so that I feel we all gave a significant contribution). Important contribution was given by the observations carried out by Tony Fairall (University of Cape Town).

In Bologna (~ 1975) I asked Riccardo Giovanelli (he was back from the States after getting his PhD) to find out the best radio telescope to survey galaxies in HI. The choice was for the Arecibo dish and we applied for time. Prof. Oort, with whom I had discussed my work in various occasions by letter and during a short visit in Leiden, encouraged me very much to pursue HI work agreeing that in addition to measuring redshift to study the LSS we could measure the HI content. Later Riccardo (Giovanelli) joined for a little while the group of Astronomy at the University of Oklahoma and we started to get data using the Arecibo dish. Since the paper by Gunn and Gott (1972)[48] and the work on the interaction between gas and galaxies in clusters (see the excellent review by Craig Sarazin (1986)[49] and references therein) I thought the best way to study the interaction was the HI since the weaker gravitational potential on the outskirt of a galaxy would ease depletion by other forces, ram pressure in particular. That the time was ripe for such work is demonstrated by the paper by Davies and Lewis (1973)[50] [of which at the time I was not aware] where for the first time is detected HI deficiency in Virgo cluster galaxies and by the extensive work carried out at the Meudon Observatory by Balkowski's group (1973[51], 1974[60]). Galen Gisler (1976 and PhD thesis)[52] simulates these interactions toward a detailed understanding of ram pressure. In figure 10 I show one of the early results where the Cluster galaxies HI deficiency is clearly detected in the Coma – A1367 super-cluster.

In the Eighties the CfA became a gold mine of detailed information and was the first step toward the beginning of the many large surveys that enabled us to understand details on the distribution and evolution of galaxies in the Universe. John Huchra, Marc Davies and Margaret Geller, that I know, were the main players. We have now millions of accurate observations, the SDSS opened the way to new knowledge in various fields, and physics and simulations show that we have a rather good understanding of the Large Scale Structure. We are now living an era in which high precision Cosmology is possible and accurate observations become a real test for theories. The progress done in the theory has been tremendous as well and it would be impossible to illustrate the whole picture, however various excellent reviews and articles can be found in the literature. In the following I will rather briefly refer to a few advancements in the field that were particularly impressive.

IV. Hunting the mass

So far we were following the light and to this end very large surveys and simulations became nowadays extremely detailed and rich of information. One of the best examples, as I said, is the perused SDSS. The Virgo simulation, as well other revealing details on the formation and evolution combining the gas hydrodynamics to the dark matter simulations, is one of the best known. On the other hand the master of the Universe is gravitation so that we must have ways to observe the distribution of mass and determine eventually whether mass and light have similar distributions.

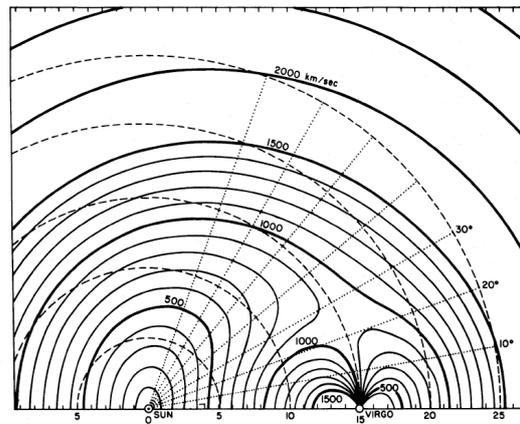

*Figure 11. Concentric dashed circles represent the unperturbed Hubble flow. However the observed distribution of mass exert a breaking force on the expansion velocity of the galaxies acting like a viscosity perturbing the regular flow. In the plot Tonry and Davis[53] use an infall velocity of 400 km/s.*

After the long silence that followed the remarks[ll] by Zwicky in 1933[54], Rood et al. (1972) evidenced the problem of the missing mass with a robust analysis of the Coma cluster; the community started to think seriously about it. Page and Rood started a spectroscopic program on clusters of galaxies (end of the sixties) that later has been carried on by Chincarini and Rood.

But the real task was that to map the distribution of mass on larger scales independently of the distribution of light. Ron Kantowski in 1969[55] published a pioneering work in which he showed how an inhomogeneity (the Coma cluster) in the distribution of matter would cause a perturbation in the Hubble flow; perturbation that clearly depends on the cosmology. Peeble (1976) developed the theory and applied it to the Local Supercluster while later on and with a large amount of data Kraan-Korteweg (1986)[56] made a very detailed analysis determining accurately the local infall motion. In between Tonry and Davis (1981)[53] did what I would call a milestone analysis depicting in a very clear way the Hubble flow perturbations expected due to the observed local distribution of galaxies. A perturbation in the density causes a perturbation in the Hubble flow and it is clearly a function of the cosmological parameters, for small perturbations we have $\Delta v / r H \approx \Delta \rho / \rho \, f(\Omega)$ while in the presence of large density fluctuations we must solve the Friedman equations. Indeed the beauty of these rather local observations is that they allow the estimate of the cosmological density parameter, we look at home and yet we estimate the dynamical parameters of the Universe. At the time Tonry and Davis, Figure 11, estimated $\Omega=0.5$ (+0.3 -0.15).

A fundamental step forward on the distribution of mass on large scales came with the work of Avishai Dekel and his collaborators. Edmund Bertschinger and Avishai Dekel (1989)[57], but see also the following papers and the review by Dekel in ARAA 1994[58], show that by reconstructing the 3D velocity field under the assumption that the smoothed field is a potential flow (perturbations that grew by gravity) it is possible to

---

[ll] Falls sich dies bewahrheiten sollte, würde sich also das üherraschende Resultat ergeben, dass dunkle Materie in sehr viel grosserer Dichte vorhanden ist als leuchtende Materie.

recover the distribution of mass, Figure 12. This is shown to be in good agreement with the distribution of light, with the presence of Voids and with previous work by Lynden-Bell et al. (1988)[61] describing the Great Attractor[mm]. These are fundamental steps toward the understanding of the Large Scale distribution and the concept of the different intensity of the fluctuations of matter and light. In addition they are leading to a full comprehension of the limits and procedure in the estimate of the fundamental cosmological parameters (see Sandage 1961)[62].

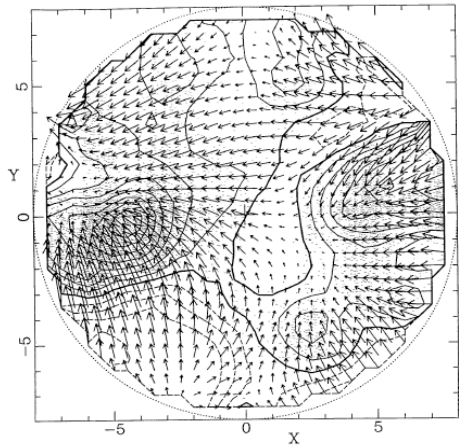

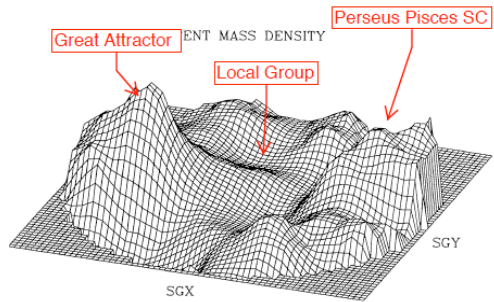

*Figure 12. Fluctuation fields of velocities and the distribution of mass-density as recovered by the analysis (POTENT) developed by Dekel and collaborators. The vectors shown are projections of the velocity field in the 3-D frame. The size on the super-galactic plane is about 200 x 200 Mpc.*

An even more powerful method, on the other hand, was on the horizon. General relativity tells us that gravitational fields bend light and acts as a lens, Figure 13. Following the article by Einstein (Science, 1936[63] – see however the paper by Renn, Sauer and Stachel Science, 275, 184, 1997[64]), Zwicky in his 1937[65] article states: "The observations of such gravitational lens effects promises to furnish us with the simplest and most accurate determination of nebular masses". Among the various later articles on the subject, Bourassa, Kantowski and Norton (1973)[66] – but see also references therein - compute the lens effect of galaxies and later on theories and observations developed tremendously. The real excitement and the interest of the community came, however,

---

[mm] Chincarini (1982)[69] in the Rio de Janeiro lectures (reprinted by the University of Oklahoma) page 26 states: "The possibility exists that the velocity field is perturbed by not yet well studied, or more distant, density fluctuations (lack or excess of matter) so that an answer can 'be given only after we know more about the density distribution. Observationally, for instance, the effect of the large cloud at R.A. = *12h54m* and D. = -*15°.2* needs to be fully evaluated. The possibility also exists that the present difficulty is due to error or to large-scale primordial vorticity of the Universe. It is therefore very important to have deep redshift surveys and understand the role played by the negative and positive perturbations". Scaramella et al. (Nature - 1989)[70] following the completion – we had the data at hand before publication - of the southern extension of the catalogue of Clusters of Galaxies (Abel, Corwin and Olowin, 1989)[71], detected a large density perturbation not too far from the Great Attractor.

with the detection and discovery by Lynds and Petrosian (1986 BAAS)[67] and by the group of Fort (Soucail et al. 1987)[68] of the gravitational lensing in clusters of galaxies
.

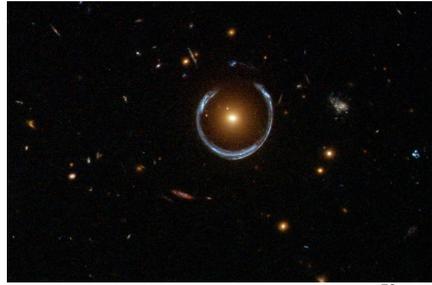

*Figure 13. The Horseshoe Einstein Ring from Hubble Space Telescope is one of the most beautiful demonstrations of gravitational lens when the source and the lens are aligned almost perfectly on the line of sight. From NASA http://apod.nasa.gov/apod/ap111221.html*

Tyson with his pioneering deep exposures, the first ever, evidenced the effect of weak gravitational lensing as well. It is due to the theoretical work and to the detailed observations and interpretation by Nick Kaiser and Squires (1993, Ap.J. 404, 441)[72] but see also Tyson et al., 1990, ApJL, 349, L1)[73], however, that the big development of this discipline started.

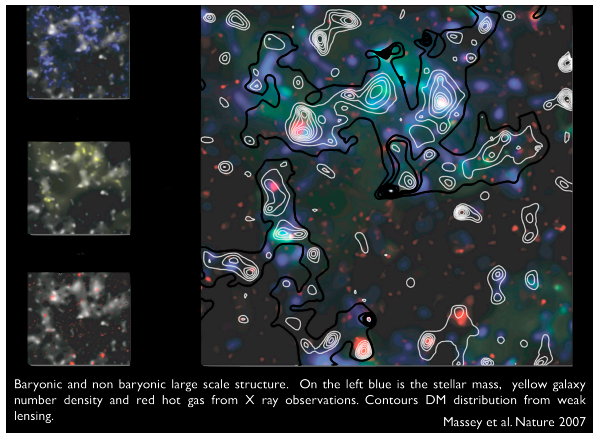

*Figure 14. On the left figures the dark matter distribution is shown as a linear grayscale. Blue in the top left corresponds to the stellar mass, yellow in the middle figure galaxy number density and red the X ray diffuse brightness (no point sources). The Figure on the right is a composition of the various components and the white contours lines the projected distribution of matter as measured from weak lensing. For quantitative details and analysis, see Massey et al. Nature 445, 286, 2007[74].*

Large -scale structures, composed mainly by dark matter, deflect by the action of gravity light rays and distort coherently the observed galaxies. The effect enable us therefore to estimate the mass of galaxies and clusters of galaxies, to detect, in brief, mass fluctuation along the line of sight and to estimate the geometry of the Universe. The idea of mapping the distribution of matter on large scales in such a clean and direct way is of paramount importance. The technical difficulties are huge due to the need a) of very high-resolution images, b) rather large fields of view and c) full control of the instrumental and cosmological systematic errors. The intrinsic alignments of galaxies generated at their formation in close proximity of the existing distribution of mass (Heavens et al. 2000[75] and Heymans et al. 2006[76] among others) and the errors related to systematic must be understood and measured in order to get an estimate of weak-lensing effect. Massey et al. (Nature 2007) combining redshift observations with weak-lensing, optical and X ray observations show the power of these observations and give a feeling of the "touch with your finger the Cosmo" effect, Figure 14 and Figure 15.

Using the redshift it is then possible to derive the space distribution of baryonic matter and dark matter, Figure 15, in a way that is completely free of astrophysical assumption and using a probe that is sensitive to matter independently of its Nature. This is the tremendous progress done in about 40 years thanks to the progress in the instrumentation, theory and dedication of many astronomers.

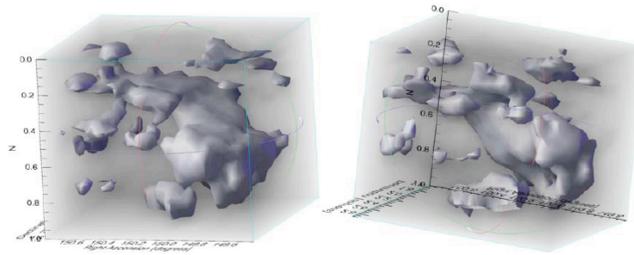

*Figure 15. Reconstruction of the 3D distribution of matter as derived by Massey et al., Nature 445, 286, 2007.*

The picture we derive is that the combination of a large number of observations using different techniques leads to a very vivid description of the Cosmo with knowledge of the distribution of the density of matter on large scales and the related perturbations to the Hubble flow.

**V. Toward a new era**

> *I congratulate you on your successful escape from the sphere of theoretical physics ….....*
> *Letter (23 April 1934) by Rutherford to Fermi*

Large structures and filaments will perturb the Hubble flow in their vicinity. A perhaps naïve approach in this direction has been attempted by Baffa et al. (1993)[77] using redshift and H-magnitudes (a method originally developed and applied by Marc Aaranson[nn]) observations. The gravitational potential of a filament can be easily approximated with a linear model of the distribution of mass and for a mean over-density of the super-cluster with $\Delta\rho/\rho = 9$ the mass over a 100Mpc has been estimated to be of about $10^{16}$ M$_\odot$; it has been found that the structure is shrinking. To some extent what we did was an original approach, very simple and limited however in scope. Kaiser (1987)[78] approaches the whole matter in a statistical way and shows that the Hubble flow is perturbed on all scales and while the cluster velocity dispersion causes an elongation[oo] on cluster scale in the redshift space (assume a spherical distribution of objects for the real space), on much larger scale the contours are kind of flattened. This approach is well depicted by the work of Peacock et al. (2001)[79] where contours of equal intensity of

---

[nn] Marc Aaranson died in 1987 at KPNO while observing. Joanna Manoussoyanaki, coauthor of the paper Baffa et al., died of cancer in 1988.

[oo] It is of interest that Jackson already in 1972 (MNRAS 156, 1P) noticed, with the few radial velocities available at the time, that galaxies appear to fall into long chains or cigar-shaped configurations. Often in the literature this elongation is called "the finger of God". I do not like to call it this way because it is historically wrong. Bart Bok looking at the HI map, see Oort et al., 1958, MNRAS 118, 379[81], noticed that near the sun the gas seems to point in long fingers radially away from the sun. Since the reconstruction is based on the assumption of circular motion, the presence of non-circular motion would considerably displace the position of the clouds and stretch them artificially along the line of sight. The comment by Bart Bok was that the phenomenon represents the fingers of God pointing at us telling: "You are wrong, you are wrong, you are wrong!" See also The Physical Universe by Frank Shu, page 273.

the two points autocorrelation function ξ(σ, π)[pp], Figure 16, are plotted as a function of the separation on the sky σ and of the separation (redshift) along the line of sight. The work by Kaiser (see also Dekel and Lahav 1999, ApJ 520,24)[80] showed that using this analysis it is also possible to compute

$$\beta \equiv \Omega_m^{\gamma \sim 0.6} / b \; ; \quad \left.\frac{\Delta\rho}{\rho}\right|_{Light \equiv galaxies} = b \left.\frac{\Delta\rho}{\rho}\right|_{Mass} \quad . \text{ For } \Omega_m = 0.27 \text{ and } \beta = 0.49 \pm 0.09$$

(Hawkins et al. 2003, MNRAS 346, 78)[135] we have b = 0.93.

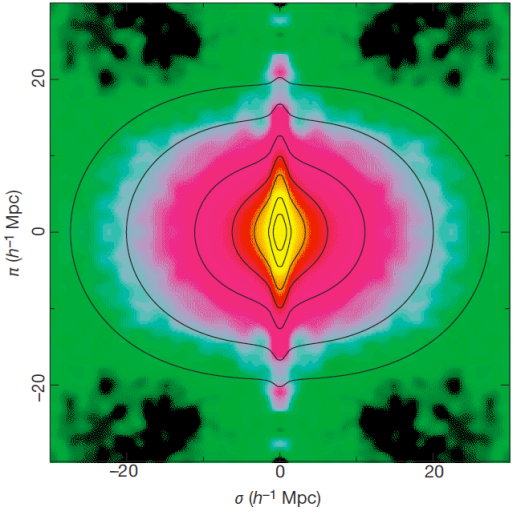

Figure 16. The two points autocorrelation function as measured by Peacock et al., 2001, using the 2dFGRS survey. Here are clearly visible the small scale elongation due to the velocity dispersion in clusters and the pumpkin like shape due to the flow perturbation on larger scales. The continuous lines are lines of constant intensity of the autocorrelation function.

The light density fluctuations follow the mass density fluctuations in spite of the fact that the clustering of galaxies is a function of the morphological type also on large scales and low-density environments, Giovanelli, Haynes & Chincarini, ApJ 300, 77, 1986[82]. Guzzo (2008)[83] estimated these parameters using a sample of galaxies at <z> = 0.77 and shows that the growth rate function $f(z) \approx \left[\Omega_m(z)\right]^\gamma$ (Linder et al. 2005)[136] is a robust way to discriminate among models on future large samples of galaxies. Peacock et al. measured the autocorrelation function to a separation of about 25 Mpc, and Guzzo et al. sample covers a Volume of about (200 Mpc)$^3$.

The Large Scale Structure is the results of primordial seed perturbation that, according to the standard theory, might form during the inflation period. It is during the radiation dominated era, when baryons and photons are coupled and share the same temperature, that such perturbation induce acoustic oscillations that are detected as Temperature fluctuations of the microwave background. These acoustic waves imprint a characteristic clustering scale in the matter we see today, Figure 17. This is a feature that is foreseen by the ΛCDM and it depends rather strongly also on the amount of baryons we have in the Universe[qq]. In short we have a new standard length (the commoving length of this standard ruler is defined by high precision MWB observations) that requires however a large amount of accurate observations.

---

[pp] See "The Large-Scale Structure" by P.J.E. Peebles, 1980, Princeton University Press.
[qq] Naturally the constraint must be in agreement, that is within the range of allowed values, with the primordial nucleo-synthesis.

The challenge of modern cosmology likely is beyond the redshifts probed by Supernovae: this may be one way to go. Weak lensing data, coupled to redshift surveys, will measure the growth of Structures. This growth depends in a direct way from the expansion rate and if this is in disagreement with what we measure using the BAO bump (the commoving length of this standard ruler is defined by high precision MWB observations) then we may have to question the theory of General Relativity.

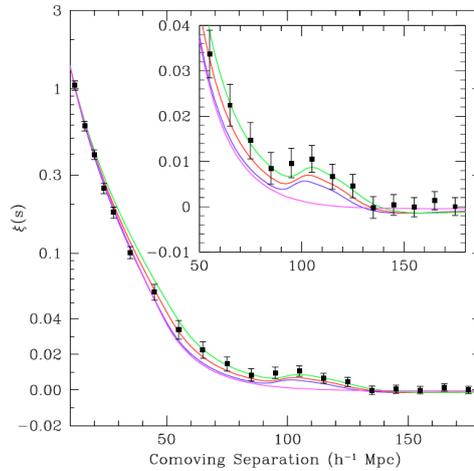

*Figure 17. The acoustic peak detected by Eisenstein et al., 2005[84], measures of the autocorrelation function based on a spectroscopic sample of 46748 luminous red galaxies from the Sloan Digital Sky Survey. For early cosmological applications see Percival et al. 2007[85].*

In a letter to Nature (1990)[86], Broadhurst et al. publish the results of four pencil surveys and detect "a remarkably regular one dimensional distribution with most galaxies lying in discrete peaks separated by 128 $h^{-1}$ Mpc". At that time we had some disagreement in the interpretation: was this a Voronoi Clustering model (van de Weygaert, 1991)[87] or a fluke (Kaiser and Peacock 1991)?[88] I figured it is interesting to mention these findings in connection with the detection of the BAO that have the same scale length.

**VI. What next?**

Perhaps the question we have to ask ourselves is whether or not we are on the right track. The achievements made on the last decades are tremendous and it seems to be only the beginning. Quantitative cosmology is flourishing and it will also in view of the new instrumentation that is being constructed. Simulations, theory and observations complement each other in an unprecedented way clarifying the formation of structures and galaxies. Dekel et al. (2009)[89] show by high-resolution LSS simulations that the merger picture is not the whole story; stream-driven accretion via the filamentary structure may lead to the formation of discs and spheroids. Streams of cold gas, sometime knotty, penetrate the shock-heated media of massive dark matter haloes and form galaxies. Governato and his collaborators in a series of papers (see in particular Governato et al., 2010[90], and Governato et al., 2012[91]) explain the formation of dwarf galaxies and how the interplay between dark and baryonic matter significantly flatten the original cusp profile for galaxies with mass $< 10^9$ solar masses. This is to say that the access to supercomputer and the very advanced simulations allow looking to the whole picture in a global way and with high resolution. Of primary importance are the visualization of these simulations as those made in various fields and on the streaming flow by Dekel and collaborators. These help the understanding of what is going on (see for instance the visualizations made by Andrew Pontzen using the simulations carried out on the Darwin supercomputer in Cambridge (UK) based on the GASOLINE Code by J. Wadsley, T. Quinn with metal cooling software written by Sijng Sheng). Elena

D'Onghia (2013)[125], with her collaborators, uses computer simulations to understand the formation of the galaxies, building blocks of the LSS. They show for the first time that stellar spiral arms are not transient features, as claimed for several decades, but they are self-perpetuating, persistent and surprisingly long lived. It seems that all together we have robust machinery based on a mosaic that is improving and becomes more and more detailed at all scales as a function of time. Are we 100% safe?

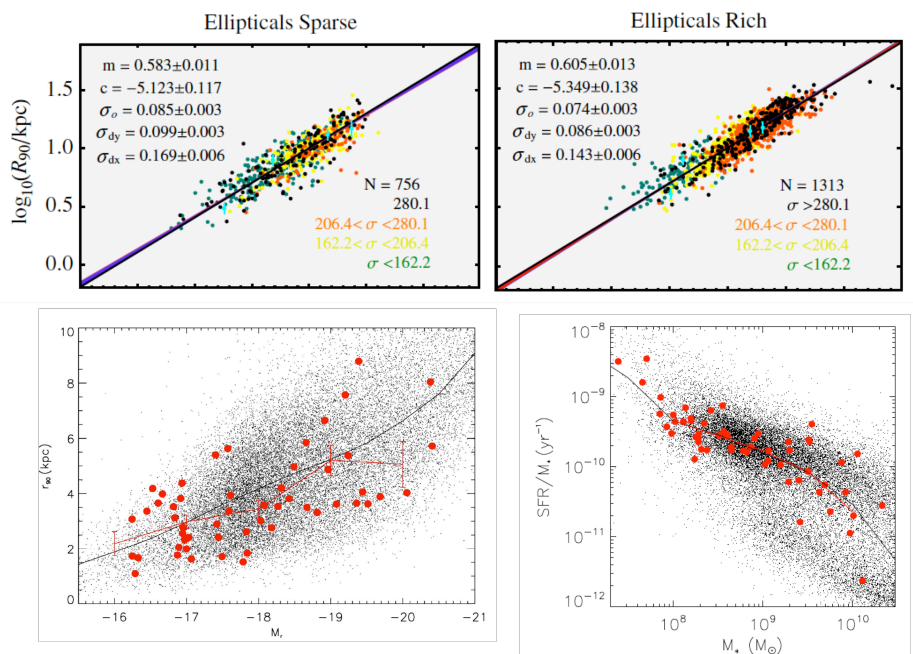

*Figure 18. Top: Nair et al. (2011) compare the Petrosian size – luminosity relation between elliptical galaxies located in rather dense regions with the relation of those that are detected in regions of low density. The size is not sensitive to the environment. Bottom: Kreckel et al. (2012) in their survey of Voids galaxies find a similar results in a plot size – Luminosity (left) and in addition show that the SSFR (Specific Star Formation Rate) – Mass relation is similar to that observed for a nearby sample of SDSS galaxies.*

Certainly the main picture is correct since so many tiles are fitting in. On the other hand Jim Peebles insisted for about two decades in looking into anomalies that may help in learning about fundamentals. To name one: the anomaly related to the number of missing galaxies in voids and the fact that galaxies seem to be unaware they are in voids, Peebles 1989[92], 2001[93], 2007[94], 2012[95] and references therein. With the assumption of a Gaussian random distribution and the fact that, according to ΛCDM and to the observations of the general field[rr], bright galaxies tend to prefer the densest regions it seems anomalous that two bright galaxies, M101 and NGC 6946, are located at the edge of the Local Void. Peebles makes the point that we should expect in the Local Void at least 10 galaxies in the magnitude range $-18 < M_B < -10$ and these are not observed. And again while it is expected that the environment influence the formation and evolution of galaxies, early type galaxies seem to be indifferent to their environment; the void

---

[rr] Giovanelli, Haynes and Chincarini, 1986, ApJ 300,77[98].

galaxies [in Bootes] seem to be unaware of the fact that they exist in a huge [Bootes] underdense region" Szomoru et al. (1996)[137].

Kreckel et al. (2011[96], 2012[97]) carried out surveys and simulations to settle the problem of galaxies in Voids; here a Void is defined as a region of the space where the density of galaxies is less than half the cosmic density. The simulations show quite clearly that there should be a rather large population of galaxies in Voids and in particular a significant population of faint galaxies with $M_r \sim 14$, the latter likely too faint to be detectable with on going surveys. Such galaxies should have a rather high Specific Star Formation Rate (SSFR) and a rather high gas content. This is not what the observations show. The galaxies detected, see for instance the KK246 with $M_{HI} = 1.05 \pm 0.08 \; 10^8 \, M_\odot$, are generally gas rich with rather normal star formation rate ($\leq$ 1-2 $M_\odot$) and not observable trend with mass of the SSFR is detected. In their sample of 60 galaxies, 41 of which detected in HI (the sensitivity is for detection of $M_{HI} \geq 3 \; 10^8 \, M_\odot$), they stress, Figure 18, that the observed galaxies do not differ in their properties from galaxies of the general field having the same luminosity. Furthermore probing a field of about 485 Mpc$^3$ they find no evidence for the missing low luminosity Voids population.

The ALFALFA survey (Arecibo), Giovanelli et al. 2005, AJ 130, 2598[99], is an excellent complement to these results and, when completed, will likely say the final word since as a blind survey detecting hydrogen down to a mass of about $10^6 \, M_\odot$ it will probe in a unbiased way a large part of the sky. The preliminary results of the Arecibo survey agree with the fact that the missing Voids galaxy population remains an unresolved problem, Saintonge et al. (2008)[100]. In addition and to further support the lack of anomalies in the building up of the galaxies detected in Voids Nair et al. (2011)[101] measuring the size, for instance, of galaxies in different environments find, Figure 18, that galaxies are not affected by the location in which they are born and evolve. However according to Tinker et al., 2008, the Peebles's voids anomaly may not be a problem at all for the ΛCDM. The big and coarse picture, in few words, shows that while galaxies tend to cluster according to the morphological type, (Giovanelli, Haynes, Chincarini 1986)[98], that is the morphology seems to be aware of the environment either during formation or evolution, the mechanism at work, once the galaxy has decided the morphology, does not care about the environment. In addition to the broad picture we likely need to better understand what mechanism limit in Voids the growth of galaxies and deplete them, at least in part, of the building blocks for galaxy formation. Only the low mass are left there and do not participate to the merging growth due perhaps to an early local sweeping of the IGM. Could highly energetic phenomena bias the ΛCDM game?

In this game where accurate (in the meaning given by Jim Peebles) and precision Cosmology complement each other toward the new step in the fundamental understanding, we are left with these three big unknowns: a Universe made of matter we do not see (DM – Appendix III), an Energy that we hardly know where is coming from and what it is (DE or quintessence where the energy density is associated with a time-dependent scalar field). And however the concurrence of the approaches we described above could be helped by the huge developments we recently had and could have in the future in the knowledge of the physics of clusters of galaxies and in their use as probes of Cosmology.

Random density fluctuations with a Gaussian distribution have a zero mean value and if negative cannot reach a value smaller than -1. Some of these are such that Δρ/<ρ> > 1.68 and follow a non – linear collapse. Smoothing such fluctuations we clearly have a smaller number of peaks on larger scales that collapse later, Figure 19. Clusters of galaxies involve in the distribution of matter lower peaks, larger scales and later collapse. Since the seminal paper by Press and Schechter, 1974[102], simulation and theoretical improvements clearly showed how powerful a tool for Cosmology are clusters of galaxies. Clusters of galaxies form rather late in the evolution of the Universe and are density enhancements that are on the limit between large scale structures that had not yet time to collapse or whose over-density will never allow the collapse and cosmic bound objects in dynamical equilibrium. They are very sensitive to the growth of linear density perturbation and their physical parameters are a function of the redshift – distance relation. Because of this both the mass function and the cluster density profile depend on the cosmological parameters and the equation of state: p=w ρ with w = -1 for a non-evolving cosmological constant in the context of the General Relativity theory.

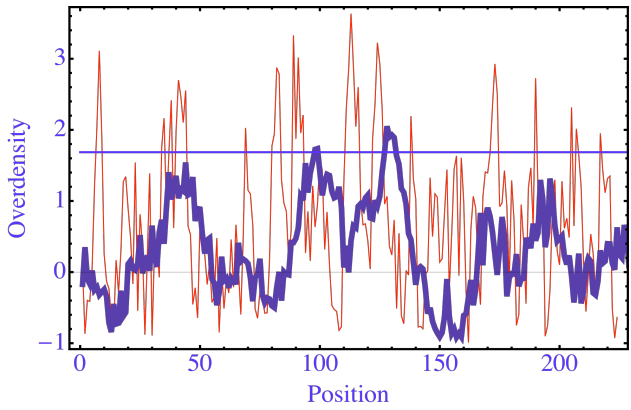

Figure 19. Red line: Random Gaussian distribution fluctuations. Blue line the same fluctuations after smoothing. A smaller number of peaks are above the critical over-density so that less structure will form at that scale and later on in time (the collapse time is proportional to $\rho^{-1/2}$.

Recent, and on going, improvements in the theory, simulations and quality and quantity of the observations make this tool, that will complement other techniques, one of the most robust for the next cosmological steps. To have an idea of the interplay of the parameters we summarize a few equations that give explicit dependence on some of the parameters involved; for details see however Hallman (2007)[103] and references therein.

Following Hallman (2007) the co-moving halo number density as a function of Mass and redshift is:

$$\frac{dn}{dM}(M,z) = -0.315 \frac{\rho_0}{M} \frac{1}{\sigma_M} \frac{d\sigma_M}{dM} \exp\left\{-\left[0.61 - Log[D(z)\sigma_M]\right]^{3.8}\right\}$$

D (z), the growth function, is a function of the cosmological parameters and of their evolution in time. The beauty of these new developments, their roots are in the visionary work of Press and Schechter (1974), is that via the rms amplitude $\sigma_M$ (normalized with the present day rich clusters fluctuations on the 8 h$^{-1}$ Mpc scale) of the density fluctuations as a function of mass at any redshift, they describe the whole story of the formation and evolution of the structures:

$$\sigma^2(M,z) = \int_0^\infty \frac{dk}{k} \frac{k^3}{2\pi^2} P(k,z) |W_R(k)|^2$$

&

$$\frac{k^3}{2\pi^2} P(k,z) = \left(\frac{c\,k}{H_0}\right)^{3+n} T^2(k) \frac{D^2(z)}{D^2(0)}$$

The function $W_R(k)$ is the Fourier transform of the real space top hat smoothing function and P(k, z) the power spectrum of the fluctuations at a given scale length. The transfer function T(k) reflects details of the growth of structures in the various scenario reflecting the interplay of DM, baryons and neutrinos during the growth of perturbations.

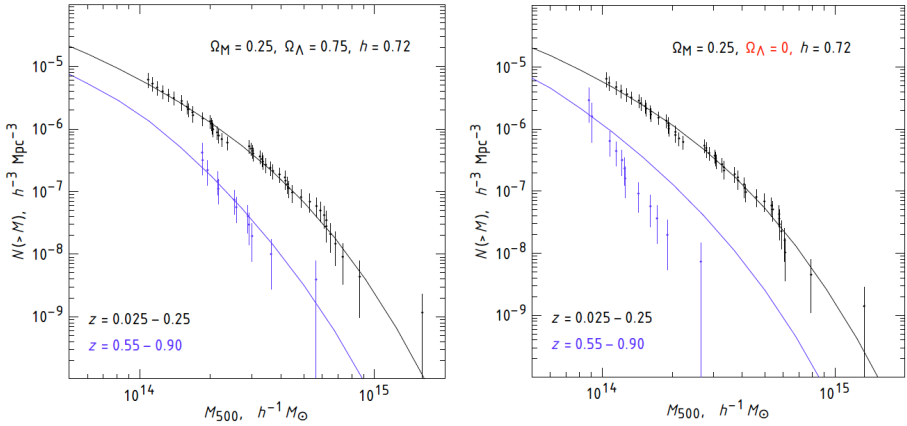

*Figure 20. These plots illustrate beautifully (Viklinin et al., 2009, ApJ 692, 1060)[105] the sensitivity of the cluster mass function to cosmology. On the left panel the fit of the observed mass function for clusters of two subsamples at different redshift with parameters satisfying "concordance" cosmology. On the right the comparison data model for a $\Omega_\Lambda = 0$ cosmology. The disagreement theory – observations is evident. The observed mass function changes because the difference in the distance – redshift relation and the models differs in their prediction of the growth of structures.*

One of the most recent studies in this direction is that by Vikhlinin et al., 2009[104], from which we use here part of their conclusions and we refer to that paper, and to Viklinin et al.,2009, ApJ 602, 1033, for details on the derivation of the relevant parameters. As shown in Figure 20 the method is quite sensitive to the cosmological models and to the model regulating the growth of structures due to the fact that the expansion of the Cosmo itself modifies the evolution. In other words clusters feel the dynamics of the Universe.

The concordance model following the results of the SNe surveys (Riess et al., 1998[106] – Perlmutter et al. 1999[107]) asks for a cosmological *constant* that differs from zero. This implies that the conservation law equation $\frac{\partial \rho}{\partial t} + \frac{\dot{a}}{a}(3\rho + 3p) = 0$ for a constant density of energy has the only solution p = -ρ. On the other hand the distance – redshift relation for SNe could be the consequence of an unknown energy evolving in time. Expressing in a general form p = w ρ we have as stated above w = -1 for a cosmological constant, w = 0 for matter and w = 1/3 for radiation. To simplify we can write the above equation in a simpler form $\frac{\partial \rho(t)}{\partial t} + H(t)(3\rho(t) + 3p \equiv 3w\rho(t)) = 0$ whose solution is: $\rho(t) = C\, Exp\left[-\frac{3}{2}H(t)t^2(1+w)\right]$. The task is the estimate of w.

Expanding on the clusters of galaxies analysis, Vikhlinin et al. derive the results depicted in Figure 21 by combining the constraints from four cosmological datasets: Baryonic acoustic oscillation (BAO), Cosmic Microwave Background (WMAP), Supernovae (SN) and Clusters of galaxies (clusters). The normalization of the amplitude of the linear perturbations at a length scale 8 h$^{-1}$ Mpc has been derived from the local mass function: $\sigma_8 = 0.813\, (\Omega_M/0.25)^{-0.47}$.

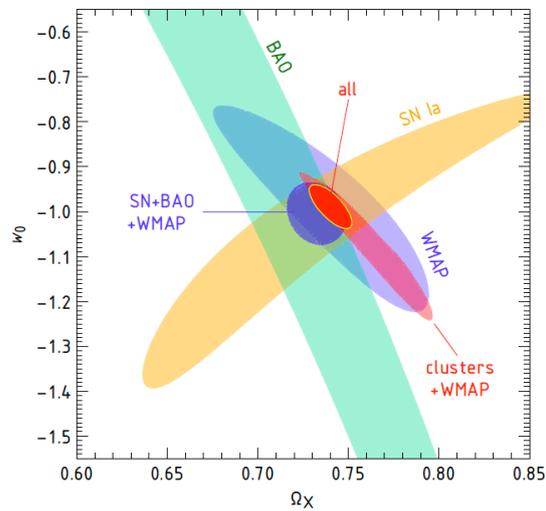

*Figure 21. Constraints of the dark energy equation of state using clusters of galaxies in combination with data from WMAP, BAO, SNe; $w_0$= -0.091 ± 0.067 (± 0.04 systematic), $\Omega_{Dark\ Energy}$ = 0.74 ± 0.012 with $\sigma_8$ (local mass function) = 0.813 $(\Omega_M/0.25)^{-0.47}$; from Viklinin et al. 2009.*

Accurate and precision cosmology tends to overcome difficulties and give quite a consistent picture. Even assuming that most, if not all, of the minor inconsistencies could be explained with the fine tuning of the equation of state, note that the estimates seem to be rather accurate, the fact remains that we have no information at all, no detection of non-baryonic matter or dark energy, about DM and DE. And this is a route that must be scouted till we find an answer. The literature is rich with the various attempt to modify the theory of General Relativity, among which the best known is probably related to modifying the gravitational force via a Yukawa interaction [MOND – Modified Newtonian Dynamics, Milgrom 1983[108] – 1984[109]), to explain "in primis " the flat rotation curve of clusters of galaxies without the need of the presence of Dark Matter. Introducing a gravitational potential of the form (Sanders 1984[110]): $U(r) = \frac{G_\infty M}{r}\left(1 + e^{-r/r_0}\right)$ the circular velocity is given by $V = \sqrt{\dfrac{G_\infty M}{r}\left[1+\alpha\left(1+\dfrac{r}{r_0}\right)e^{-r/r_0}\right]}$ and the flat rotation curves could be fitted without invoking the presence of a Dark Matter halo. For an exhaustive discussion of the observations and related theoretical limits see Sanders 2010[111] (The Dark Matter Problem, Cambridge University Press). The point is that a modification of the General Relativity demands a minor correction on very large scales (we do not modify on scales of the solar system) or, equivalently, in regions of very low matter density.

Clusters, as we mentioned earlier, offer another parameter that is sensitive to cosmology, the cluster profile. The cluster parameters and their values as a function of radius are all sensitive one-way or the other to gravity and the profile must reflect this dependence. To do this we need large surveys and accurate measures of cluster parameters.

A recent and very interesting approach to this issue has been attempted by Terukina and collaborators based on the Chamaleon theory (see MOND's potential). The main characteristics in these theories is the presence of a scalar field which couple to ordinary matter density and leads to the presence of a new Yukawa interaction so that the gravitational force is written as $F_{1,2} = \dfrac{G\, m_1 m_2}{r^2}\left(1+\alpha_1\alpha_2\, e^{-m r}\right)$ and the scalar fields when coupled to matter have indeed a matter dependent effective potential. Newtonian gravity is recovered in high-density region.

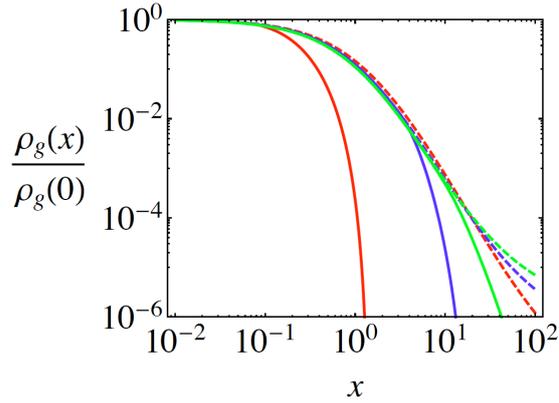

*Figure 22. Cluster profile of the gas distribution as detected in the X-ray as a function of the distance from the center normalized to a scale that is characteristic of the halo. Green, Blue and Red refer to masses $4\,10^{14}\,M_\odot$, $10^{14}\,M_\odot$, $4\,10^{13}\,M_\odot$ respectively. The continuous lines are computed with the presence of the Chamaleon force and the dashed lines without.*

Figure 22, reproduced from Terukina and Yamamoto, 2012[112], shows in an impressive way that cluster profiles are very sensitive to the theory. This is an important route that should also be pursued by measuring detailed physical parameters on a large number of clusters.

A cluster deep survey over a large area is, therefore, needed and would be scientifically very rewarding. Efforts in this direction started in about 1990 when Riccardo Giacconi, at the time Director of the Space Telescope Science Institute in Baltimore, was dwelling with the idea of designing a wide field X ray Telescope in collaboration with Richard Burg and Christopher Burrows (1992)[113]. The idea was very simple and applied long ago to optical telescopes by Ritchey – Chretien: by modifying the shape of the secondary mirror we can design a Cassegrain optics corrected on a rather large field of view. The design using grazing incidence optics seemed to be rather complicated and however one day I received a telephone call from Riccardo during which he mentioned they were successful[ss] in designing the optics and he was offering a survey mission to carry out in collaboration. At OAB in Merate we had the capability to experiment with thin and cheap X ray optics thanks to the capabilities of Oberto Citterio with whom I was forming the X ray group at the Observatory and the collaboration with MediaLario in the nearby City of Lecco. Our other asset that fully developed and improved later on the design was Paolo Conconi. From the first idea of making a set of small telescopes we switched to a design of a single larger telescope (this decision was taken during a meeting we had in Colorado) and we started to make design, plans and proposals. Giacconi was inspiring and following all details [including extremely interesting contacts with Space Agencies and rockets manufacturers] so that we started to build prototypes. We did not publish much at the time because so busy in the real work, on the other hand we decided to present some results at the Postdam meeting organized by Gunther Hasinger, Chincarini et al. (1998)[114]. *This publication*, that reflects the work we did during the first few years of this project, is fundamental since it *shows for the first time that the optics could be actually made and was not only a theoretical exercise*.

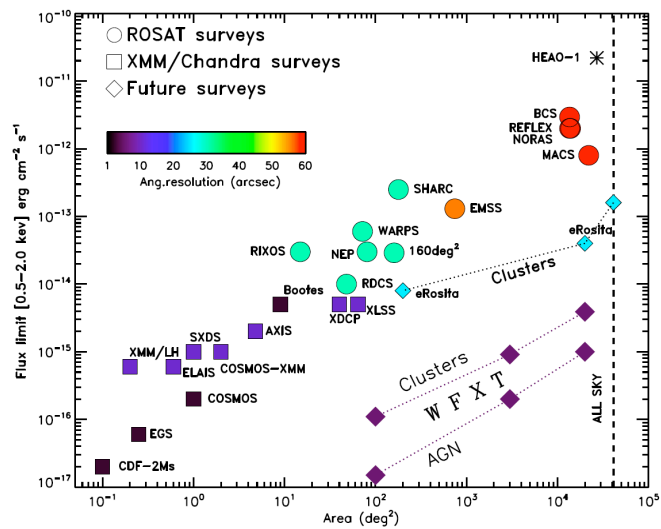

*Figure 23. A summary of past, and planned, X ray surveys of clusters of galaxies prepared by Vilkhinin. The WFXT limits are computed for a collecting area 8000 cm$^2$ and an angular resolution of 5" (HWE) and 1 deg$^2$ FOV. This resolution is hard to get unless new technology is used, see CfA SMART.*

Unfortunately it did not fly yet since, and for different reasons, NASA or ASI did not select it. The difficulties, it depends on the technology used, consist in the best compromise between resolution and weight in a survey that should reach the sensitivity of at least the X ray background. The main characteristics of the various surveys, Figure 23, have been clearly depicted by Vikhlinin et al. (2009b) where they also discuss the details of the survey that in brief would consists in the detection of a huge number of clusters and high accuracy data. The best way to go, about 20 years after we developed the wide field optics technology, is what has been proposed for the SMART mission. Here again the X ray technology follows a concept that has been very satisfactory applied at optical wavelengths, NTT and VLT: active optics. Again the difficulties with the coaxial X ray optical shells are enormous, but the idea to accomplish such a fascinating task fundamental. In this case the actuators are piezoelectric device

---

[ss] Soon after we discovered that Werner had a similar idea for the design of the wide field X-ray optics in 1977. This does not detract anything from the independent work of Burrows et al. (1992) since what counted were the science motivation, the guidance and enthusiasm of Riccardo Giacconi.

(and we refer to http://pcos.gsfc.nasa.gov/studies/rfi/Vikhlinin-Alexey-RFI.pdf for details) and hopefully a new era for clusters of galaxies and cosmology will start.

Meanwhile the understanding and observing capabilities of the Sunyaev Zeldovich (SZ) effect developed tremendously. Looking at the microwave background radiation through plasma, as it happens observing clusters of galaxies, the microwave spectrum is distorted. Photons, passing through the plasma at the temperature of about 10 keV, will be inverse Compton scattered. The percentages of photons that are scattered receive a boost that is proportional to $k_B T_e / m_e c^2$ and the spectrum is distorted because photons transit from lower to higher frequencies. The thermal effect modifies the background black body spectrum of a very small amount. For $h\nu = kT$, that is at frequency of about 56 GHz for T~10 keV, the effect is about:

$$\Delta T = 2T \int_{-\infty}^{+\infty} \sigma_T n_e(r) \frac{k_B T_e}{m_e c^2} dr \sim 0.5\, mK$$

for a bright cluster. Details as a function of frequency in Figure 24, where to evidence both the background spectrum and the distortion due to the presence of the cluster's plasma the two curves have a different normalization: the Black Body peak is a factor $10^5$ more intense than the peak of the perturbation.

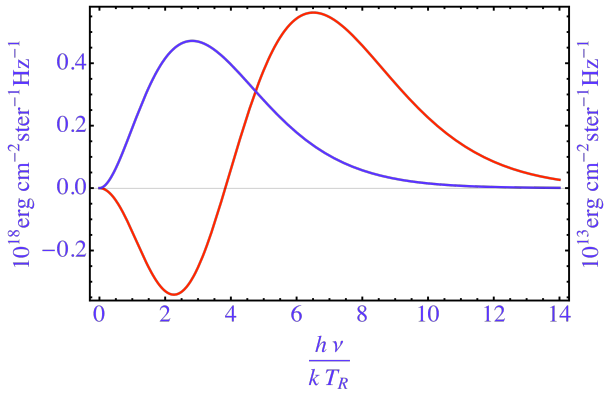

*Figure 24. The blue line is the Black Body at $T_R$ = 2.7 Kelvin. The relevant label is to the right of the frame. The red line represents the SZ spectrum distortion and the related label is on the left of the frame. The SZ effect has been computed for a rather low mass cluster: y = $10^{-5}$ (see the text for details).*

The strength of the distortion is clearly related to the number of scatters and therefore to the optical depth of the plasma. In a more accurate way it is strongly dependent from the y Compton parameter: $y = \int n_e(r) \frac{k_B T_e}{m_e c^2} \sigma_T dr$. The y parameter is related to the mass of the cluster. The distortion $\frac{\Delta T_{SZ}}{T_{MWB}} = f\left(\frac{h\nu}{kT}\right) y$ is independent of redshift!

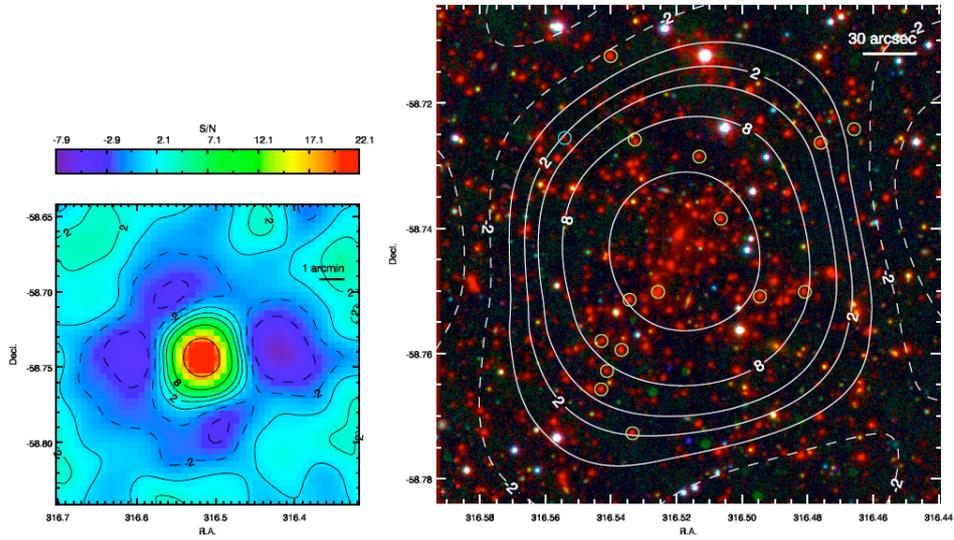

*Figure 25. This is the most massive cluster known detected using the South Pole Telescope at z = 1.132. The SZ effect contour lines are (left) superimposed to the optical image (see Foley et al. (2011)[115] & Williamson et al., 2011[116]).*

The observational field developed tremendously in the last years thanks to new facilities that are tackling this type of observations both using single dish or interferometer (first detection by Jones et al., 1993[117], using the Cambridge 5 km telescope). Figure 25, observations from the South Pole Telescope survey, shows the high resolution and the possibility of mapping the distribution of the plasma; circled galaxies are spectroscopically confirmed members. The Coupling of these observations with the X ray data, Figure 26, gives the possibility to estimate cluster physical parameters independently and high accuracy.

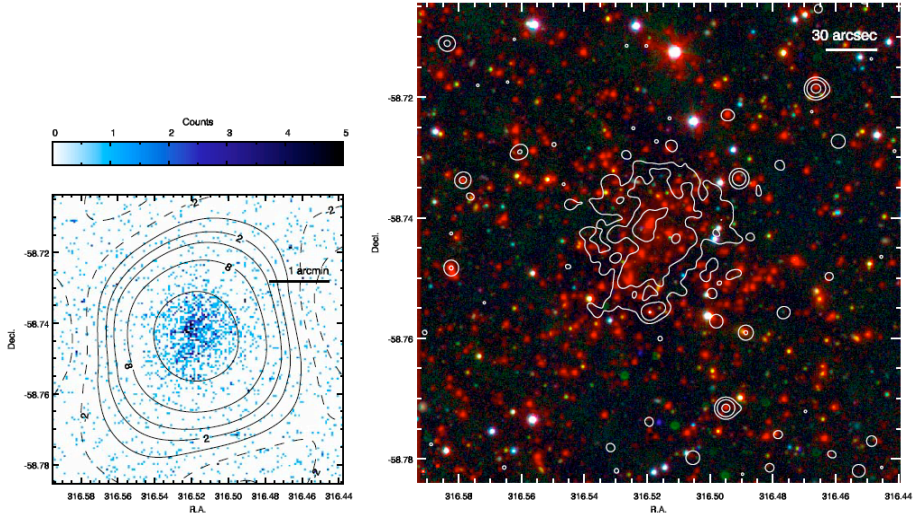

*Figure 26. (Left) Chandra X-ray image with black lines representing the SZ significance derived in Figure 23. (Right) Chandra X-ray contours superimposed to the optical/Spitzer/IRAC image (details in Foley et al. 2011).*

To mention the mass estimate of the cluster, certainly one of the most important parameters, this is derived in three different ways SZ: X-ray and optical velocity dispersion to a rather large radius, to an over-density $(\Delta\rho/\rho_c)$ = 200 and 500. The SZ image, the X-ray and the optical distribution of galaxies (the eye after-all is an excellent instrument) all show an elongation that goes about the direction SE – NW, a good way to measure asymmetries in the cluster profile. Another fascinating example is the cluster observed by Korngut et al. (2011)[118] with the MUSTANG receiver at the focus of the Green bank Telescope. Here the coupling of the SZ data with the observations by Chandra allows the study of a weak shock with the estimate of the shock velocity, Figure 25. Large surveys and detailed follow up are highly desirable and in general it could go both ways, radio surveys to be followed up by X-ray pointing and vice-versa as to eliminate as much as possible observational bias. Planck of course generates a SZ survey as well with its limitation however, since small distant clusters may be lost due to the resolution of the instrument. According to Chamballu et al., 2012[119], most of the Planck clusters should be at z < 1 in a X-ray flux range $10^{-12} - 10^{-13}$ erg/s/cm$^2$ and with T >6 keV, more or less in agreement with the sample discussed by Ade et al, 2013[120]. On the other hand Planck has a FWHM at 150 GHz of 5' so that for clusters at z > 0.5 with a diameter < 1' the sensitivity decreases. The path has been marked and, a fascinating long and rewarding way to go, outlined.

As shown above one of the parameters we can measure with observations of the SZ effect is the density once we know the temperature of the gas, a value that we easily derive from the X-ray observations. Electromagnetic radiation in its journey through the intergalactic space rotates its plane of polarization. The relation: $\theta = RM\,\lambda^2$ where $RM = \dfrac{e^3}{2\pi m_e^2 c^4}\int_0^d n_e(s)B_\parallel\,ds$ relates the rotation of the plane to the wavelength so that with measures in two (or more to avoid some degeneracy) pass-bands and the knowledge of the density we can measure the magnetic field. Not straight forward since we may have a delicate resolution matching to worry about but feasible and with LOFAR and other facilities coming into action we have opened the door to a fundamental problem, the estimate of magnetic field in cosmic objects and in large structures. Magnetic fields are poorly known and hard to measure, their intensity in the Intergalactic Medium unknown and the origin a completely open issue.

Magnetic fields interests cosmic objects at any scale and the Universe as a whole, their intensity goes from that estimated in some neutron stars, for magnetars the field is likely about $10^{15}$ Gauss to, may be, extremely low fields in the Intergalactic medium and of the order of $> 10^{-16}$ Gauss, Neronov and Vovk (2010); see however the analysis by Arlen and Vassiliev (2012) who support a zero intergalactic magnetic field.

While the matter is debated the topics is fascinating: the Magnetic Field Large Scale Structure (MFLSS) and its creation and evolution.

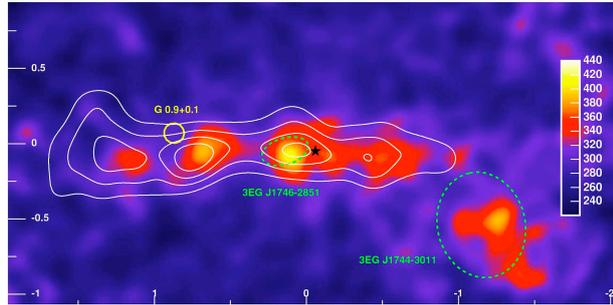

*Figure 27. VHE gamma ray image of the Galactic Center. The black star mark the position of Sgr A. the white contours indicates the density of molecular gas. Green ellipses the 95% confidence region of the Egret sources. From Aharonian et al. 2006).*

Since the discovery of cosmic rays it is well known that very high-energy gamma rays and particles exist in the Universe. An impressive result of a field that is in its full development nowadays and that shows this fact is the HESS image, Figure 27, of the center of the Milky Way. What is not known is the mechanism of particle acceleration and it is believed that magnetic reconnection and shocks may have a fundamental role in such problem. Indeed solar flares may tell us how important mechanisms of acceleration these may be. Furthermore gamma rays and particles may be good probes for the weak magnetic fields of the Cosmo.

A high energy gamma ray, E > TeV, during its journey, the mean free path is of the about $\lambda_\gamma = 80 \left( E_{\gamma 0} / 10\, TeV \right)^{-1} Mpc$, through the intergalactic medium interacting with the EBL (Extragalactic Background Light) produces pairs that loose energy via inverse Compton scattering with the microwave photons. The EBL therefore leaves an imprint in the spectra of distant cosmic sources of high-energy photons. The magnetic fields on the other end tend to diffuse somewhat the particles since the component of the velocity that is perpendicular to the magnetic field cause a gyration of the particle $Larmor\ radius = \dfrac{m_e c \sqrt{\gamma_0^2 - 1}}{eB}$. The two pairs with energy $E_e$ of about ½ the gamma ray photons IC scatters on the MWB, these photons have an energy of about 6 $10^{-4}$ eV, and loose energy on scale $D_e \sim 10^{23}$ ($E_e$/10 TeV)$^{-1}$ cm. The deflection angle is inversely proportional to the Larmor radius so that the extension of the cascade is larger at low energies. By comparing the Point Spread Function of the telescope with the expected extension of the cascade even in case of no detection it is possible to put constraints on the low energy tail of the cascade. The original high-energy spectrum of the source, on the other hand, is clearly absorbed and coupling the data analysis to simulations it is possible to set limits on the IGMF (Intergalactic Magnetic Field) as done by Neronov and Vovk (2010; for more details see also Neronov and Semikov (2009).

Magnetic fields in high density fields as those we have in the shocks observed in GRBs, SNe, AGN and other cosmic objects may form and grow from plasma instabilities as shown by simulations (see among others Nikishawa Zhang and Mac Fadyen). For the Intergalactic Medium we do not have much information and the low density of the plasma, after re-ionization, may make these processes rather difficult. On the other hand not only we may have very weak primordial seeds that some how grow during the evolution of the Universe but also the growth of instabilities and the formation of magnetic fields for a period just before recombination.

**VII. Conclusions**

When I started to get interested in extragalactic astronomy I had a few reference points. Zwicky for SNe and compact galaxies and his interest in the size of clusters of galaxies, Alan Sandage and Gerard de Vaucouleur who were debating for the value of the Hubble constant, my supervisor Merle Walker who showed to me how to research Nature. Very naively, and with a good dose of ignorance in cosmology, when I started to work on clusters with Thornton Page and Herb Rood and after talking to Zwicky, I wondered about the distribution of galaxies even if Gerard would say that cosmology comes late in life. At that time I was probably on of the very few who had knowledge of the new instruments [Roger Lynds however set up KPNO for this type of work] and when the occasion came to collaborate with Rood it seemed we were following a planned path. It is shocking to have witnessed the development that occurred from the late sixties to now! We are now moving in the sea of the Cosmo with knowledge and detailed maps so that we can approach formation and evolution and try to map the details of the geometry. The big flow of the understanding the distribution of mass and light in the Universe became a tool and a consequence of the understanding its geometry and content. In other words the details of the LSS are intimately connected to the big cosmological questions: Dark Matter, Dark Energy, Quintessence, theory of the general relativity. Who could even slightly imagine reading the perused paper written by Sandage in 1961, ApJ 133,355 "A decision between the possible classes of models therefore cannot be made until observations are pushed near to, or perhaps even beyond, the telescopic limit of the 200 inch." that we would get so far in detailing cosmology and in posing such fundamental challenges to physics and cosmology.
… Perhaps …. Cogito ergo sum, ego sum ergo Mundus est.

**Acknowledgments.**

I learned a lot from many of the scientists I met, I thank them all. I had the pleasure during my work to understand their work, respect and admire their capability in describing physics and deep concepts. I also gained from students who kept challenging with new concepts and ideas. Leonida Rosino, my teacher at the University of Padua, gave to me full support during my first steps. Merle F. Walker, my supervisor at Lick Observatory Mt Hamilton, introduced me to the real science and gave me all the tools a scientist needs. Thornton Page opened my mind to the cluster survey and introduced me to space research. With Herb Rood, a dear friend, I lived for many years in intellectual symbiosis and we were as one scientist while Jim Peebles has been, and is, for me the reference point in cosmology.  I am also grateful to all younger collaborators that tremendously enriched my mind and my life; most of them are now famous and well-known scientists. I am particularly grateful to the University of Oklahoma faculty, Physics department, for the discussions and help I got during the early seventies on most of the topics I discussed. In particular I thank David Branch; with him I often discussed my work and listened to his. Zwicky of course has been one of mine lighthouses and it has been a pleasure to listen his ideas and being challenged by his questions. Since the very beginning he told me " You may be right or wrong but you must have ideas on the Cosmo you observe, pursue and verify them".

On a more specific basis I thank for suggestions and material related to this paper Jim Peebles (again), Avishai Dekel, Fabio Governato, Hyrania Peiris and Richard Massey. Remo Ruffini for giving me the opportunity to go back in time.

**APPENDIX I.**

Following the advice of Zwicky, Rudnicki, Dworak and Flin with the collaboration of Baranowski and Sendrakowski approached the difficult and time consuming project of mapping in three colors (Blue, Yellow and Red) a region of the sky centered at RA (2000)=$11^h19^m$ and D (2000)=$35^o$ 53' that corresponds roughly to the Field 185 of the catalogue of galaxies and clusters of galaxies by Zwicky.

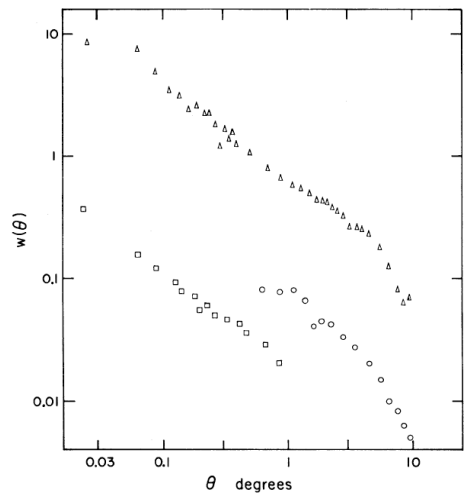

*Figure AI_1. The plot shows the two points angular correlation function for the Jagellonian catalogue (squares), the Shane – Wirtanen catalogue (circles) and the Zwicky catalogue (triangles); Peebles (1975). The function has the same slope in catalogues of different areas and depth.*

The photographic plates obtained with the 48" Palomar Schmidt Telescope, reached in the blue the $20^{th} – 21^{st}$ magnitude and the polish scientists measured 15650 galaxies. Peebles (1975) analysis of this catalogue shows excellent agreement of the autocorrelation function estimated from the Shane and Wirtanen and Zwicky catalogue. The Jagellonian field had been selected with the aim of providing a statistically uniform sample to estimate the distribution of galaxies. The field contains only a handful of bright galaxies and however it is rich of clusters of galaxies some of which are very distant.

This accurate and detailed work published in the "acta cosmologica" for some reasons did not play the role it could have at that time. In the ADS it has been referenced zero times (however we know this is not a complete statistics) and it wasn't used much (the Zwicky catalogue had been perused) by observers as a list of galaxies for further photometric and spectroscopic studies (very hard to get spectroscopy of faint galaxies at the time).

**APPENDIX II.**

My thinking about voids started in a way that motivated my surprise and curiosity with the observations, as I said, discussed in the paper written by Chincarini and Martins (1975). The paper was submitted in 1974 and the observations carried out in 1973. Obviously if we consider that redshifts are segregate in interval of redshifts it means in other redshift ranges we do not observe galaxies. Pippo Vaina after he moved to CfA invited me for a seminar that I gave at lunch, brown bag talk (as I mentioned in that occasion I had the pleasure to meet Bruno Rossi). One of the questions I was asked when I showed the observed distribution of galaxies in redshift space (empty regions and elongated agglomerates) was about the statistical evidence (I do not recall who asked). My answer was that you see it clearly by eye but I started to think; I realized that what was needed, especially for the holes (voids), was some statistical evidence. Later I gave a talk in Italy at a galaxy workshop organized by Franco Pacini at the Accademia dei Lincei in 1976[tt].

Franco never published the proceedings. After my talk Martin Rees asked whether the distribution of galaxies in redshift space, the peak characterizing the distribution of redshift in the region of the Coma – A1367 Supercluster, could be due to the effect of the limit in magnitude of the sample since more or less the distribution would peak in that region. That started to bug me even more, but my analysis wasn't yet completely ready. I had all my thinking and computations done before the Tallin meeting, and at the end of my talk I mentioned the importance of the gaps (Page 264 of the proceedings) and as a comment to Bill Tifft talk I mention my preliminary results (Page 268 of the proceedings). I also noticed at the time of writing these proceedings that at the Tallin conference Silk asks to Tifft a question similar to what Martin Rees asked me in Rome). The fact is I was not very happy with the way I derived things before going to Tallin and even Herb did not pay much attention to this part of the work. I finally improved somewhat the text and submitted to Nature also because I did not want to be scooped. I was very lucky to get Michel Fall as a referee and he made my original coarse derivation more elegant.

I consider this paper very significant and yet it got only 23 citations. It has been ignored and for some reason I discovered it was not known by many, perhaps the time was not ripe and, at that time, I could not participate to many meetings. It may also reflect a different way to evaluate work. I noticed that often the work by Rood and myself was

---

[tt] To make sure about the date (see also the reference in the talk I gave in Tallin, in the proceedings it appears as Tarenghi et al. since Massimo was a young collaborator) I asked the secretary of the Accademia di Lincei, Ms. Daniela Volpato, whom I thank very much for the quick answer. It was a workshop by the title " Galaxies and the Intergalactic medium" at the Accademia dei Lincei from the 20$^{th}$ to the 22$^{nd}$ of May 1976. The group in the following days went to Frascati to continue the presentations and discussions at the Laboratory of Space Astrophysics. I gave my talk on Friday May 21$^{st}$ at 9:30.

simply disregarded by some, perhaps as part of larger strategies or unawareness. In the review article by Alan Dressler (Nature 350, 391, 1991), for instance, the beginning of spectroscopic surveys (and of new detectors and instruments) of the seventies has been overlooked [even papers published in the journal Nature, without giving a single reference]. Tully (1982) refers to the late work on Hercules, Tarenghi et al. (1979), the work I presented at the Tallin meeting, w/o mentioning previous discoveries, and so on. However, and more important than the references, Brent refers very clearly to the picture that is emerging from the studies of the distribution of matter at large redshifts, superclusters and voids. The community was gradually assimilating the new scenario! Perhaps this is a natural process and I mentioned it since friends, no problem however, brought it to my attention. It may be that some papers, or authors and/or Institutions, are more fashionable than others.

In conclusion we were fully aware at that time about the way things were on the sky, the picture was delineating clearly. Later it was difficult to get time at optical telescopes (KPNO was the main facility for us at the time), since others became interested in this business, so that it was time to open a new way using Arecibo. I contacted Prof. Oort to illustrate and discuss [his answer was: … great, you can also measure the hydrogen content …] my plans and afterwards I asked Riccardo Giovanelli (we met when I was in Bologna) to analyze which radio telescope would be the best to carry out the research I had in mind. And the Arecibo work on galaxies, Large Scale Structure and hydrogen depletion in clusters, then started at the Arecibo Observatory (Cornell University).

**APPENDIX III.**

The understanding of DM and the detection of the related high-energy particles is one of the big questions we have in physics and cosmology, and it is beyond the scope of these proceedings. On the other hand it is worth mentioning the tight correlation between the high energy particles with small cross section we are looking for in the CDM cosmology and the possibility to have particles with higher cross section and capable therefore of generating a thermal profile of DM in the center of clusters and galaxies (and therefore of the baryonic matter that mimic the distribution of DM). Marchesini et al., 2002, D'Onghia et al., 2003 – 2004) among others, arrived at a possible estimate of a soft DM cross section. However the evidence is yet too scanty to say anything and perhaps the matter is not yet completely settled even if there is consensus among most astronomers that the profile follows the Einasto (1965 – see also Retana-Montenegro et al. 2012 and references therein) cold dark matter (no particle interaction) analytical form. For DE we must find ways to detect it and measure it, new ideas may lead to do that.

**References.**